\documentclass[aps,prx,twocolumn,reprint]{revtex4-1}
\usepackage{amsmath,amssymb,graphicx,hyperref,color}
\newcommand{\eqn}[1]{\begin{eqnarray} #1 \end{eqnarray}}
\newcommand {\tbf}[1] {\textbf{#1}}
\newcommand {\tit}[1] {\textit{#1}}

\newcommand {\trm}[1] {\textrm{#1}}
\newcommand {\densop}[2] {| #2 \rangle \langle #1 | } 
\newcommand {\ket}[1] {|#1 \rangle}

\newcommand {\LL} {\perp}
\newcounter{defin}
\newcommand{\defin}[1]{\refstepcounter{defin}\label{#1}}
\newcounter{assum}
\newcommand{\assum}[1]{\refstepcounter{assum}\label{#1}}
\newcounter{thrm}
\newcommand{\thrm}[1]{\refstepcounter{thrm}\label{#1}}
\newcounter{exam}
\newcommand{\exam}[1]{\refstepcounter{exam}\label{#1}}

\definecolor{RED}{rgb}{.8,0,0} 
\definecolor{BLACK}{rgb}{0,0,0}

\begin{document}

\title{A graph-separation theorem for quantum causal models}

\author{Jacques Pienaar}
 \email{jacques.pienaar@univie.ac.at}
 \affiliation{
 Faculty of Physics, University of Vienna, Boltzmanngasse 5, A-1090 Vienna, Austria.
}
 \affiliation{
 Institute of Quantum Optics and Quantum Information,
Austrian Academy of Sciences, Boltzmanngasse 3, A-1090 Vienna, Austria.
}

\author{\v{C}aslav Brukner}
 \affiliation{
  Faculty of Physics, University of Vienna, Boltzmanngasse 5, A-1090 Vienna, Austria.
}
 \affiliation{
 Institute of Quantum Optics and Quantum Information,
Austrian Academy of Sciences, Boltzmanngasse 3, A-1090 Vienna, Austria.
}

\begin{abstract}

{A causal model is an abstract representation of a physical system as a directed acyclic graph (DAG), where the statistical dependencies are encoded using a graphical criterion called `d-separation'. Recent work by Wood \& Spekkens shows that causal models cannot, in general, provide a faithful representation of quantum systems. Since d-separation encodes a form of Reichenbach's Common Cause Principle (RCCP), whose validity is questionable in quantum mechanics, we propose a generalised graph separation rule that does not assume the RCCP. We prove that the new rule faithfully captures the statistical dependencies between observables in a quantum network, encoded as a DAG, and reduces to d-separation in a classical limit. We note that the resulting model is still unable to give a faithful representation of correlations stronger than quantum mechanics, such as the Popescu-Rorlich box.}
\end{abstract}

\maketitle

\section{Introduction}

An essential problem faced by any scientist trying to make sense of the world is this: how do we infer causal relationships between the observed quantities, based only on information about their statistical dependencies? This problem is well known to statisticians and researchers working on Artificial Intelligence (AI), who have developed \tit{causal models} as a tool for making causal inferences from a set of observed correlations. In most practical situations, the task is made easier by the availability of additional information and physical intuition. For example, in considering possible explanations for the observed correlation between smoking and cancer, we might consider it plausible that the two are independently caused by a common genetic factor, but few people would advocate the idea that having cancer causes people to smoke -- not least because smoking tends to precede the onset of cancer, and we know that an effect cannot precede its cause. If we are simply told that two abstract variables X and Y have correlated values, the task is much more difficult. Such situations arise in theoretical work where one aims to relax the existing framework and construct more general models, or in practical applications like programming an AI to make causal inferences about data that it acquires. 

In a causal model, defined in Sec \ref{Sec:CCM}, the random variables of interest are represented by nodes and causal influences between them are represented by lines with arrows, called directed edges. The laws of physics require that no effect can be its own cause, leading to the requirement that the graph be acyclic (i.e. free of directed loops). The resulting directed acyclic graph (DAG) provides a computationally useful tool for extracting information about the statistical relationships of variables. In particular, it allows us to determine whether one set of variables is independent of any other set, conditional on the values of a third set. This information can be obtained directly from the graph using a simple algorithm, based on a concept called \tit{d-separation}. Two sets of variables will be independent conditional on a third set of variables if and only if they are d-separated by the third set in the graph.

The proof that d-separation allows one to extract all (and only) correct information about the dependencies of the variables makes causal models particularly powerful tools for representing physical systems. Indeed, we are tempted to interpret the causal structure represented by the graph as ``out there in the world" in the same sense as we can take the classical space-time manifold (which encodes causal relations between events) to be an independent and objectively defined structure. However, the program faces significant conceptual difficulties when one attempts to apply it to quantum physics. In principle, any observed probability distribution can be explained by some causal model, if we allow the possibility of hidden variables. However, as first clearly articulated by Bell \cite{BELL}, hidden-variable accounts of quantum mechanics can be challenged because they imply highly nonlocal behaviour of the model. This feature manifests itself in causal models in the form of \tit{fine-tuning}, where one is forced to posit the existence of causal effects between variables whose statistics are independent. The fact that causal models of quantum systems require fine-tuning was recently shown by Wood \& Spekkens \cite{WOOD}.

These considerations revive an old question: what does causality really mean in the context of quantum mechanics? Do we accept that there exist nonlocal hidden variables whose direct influence is in principle unobservable at the statistical level? Or could it be that the classical concept of causality does not extend to quantum systems, and that we need a completely new way of determining whether two quantum events are causally related? Following the latter point of view, we define a causal model based on quantum networks and use it to derive a graph separation rule analagous to d-separation, for obtaining the conditional independence relations between variables. Our approach differs from previous work that assigns quantum amplitudes to the nodes in the DAG \cite{TUC}, or that aims to replace the conditional probabilities at the nodes with some appropriate quantum analog \cite{LEIF13}. Instead, we retain classical probability theory, but seek a physically motivated graphical representation of the causal structure that gives rise to the probability distributions predicted by quantum mechanics. Our approach is more closely aligned with previous work in which quantum network diagrams are used to obtain joint probabilities obeying standard probability theory \cite{COE09,ORE12,CHI,HAR,FRITZ,LAS,ORE14}. Particularly relevant is the recent work by Fritz \cite{FRITZ}, in which a DAG representation of quantum correlations is proposed that encompasses our concept of a quantum causal model as will be discussed in Sec. \ref{Sec:QCMs}. Our work takes the additional step of defining a specific representation and a graph separation rule within this framework.

Recently, another DAG representation for general networks was proposed by Henson, Lal and Pusey \cite{HLP} in which d-separation continues to hold between the observed variables representing classical data. This is achieved by adding extra nodes to the graph representing `unobserved' variables, which ensure that the restriction of the CI relations to just the observed nodes produces the conditional independencies expected of a quantum network (or generalised probabilistic theory). Our approach differs from these authors, in that we consider all nodes to be in principle observable; this leads us instead to modify the criterion for obtaining the CI relations from the graph (see Sec. \ref{Sec:qsep}). The comparison to Ref. \cite{HLP} will be discussed further in Sec. \ref{Sec:Conc}.    

The paper is organised as follows: in Sec. \ref{Sec:CCM} we give a review of the relevant concepts concerning classical causal models and their graphical representation by DAGs. We include a discussion of the physical motivation for these models, and the meaning of the result in Ref. \cite{WOOD} that such models cannot faithfully represent quantum correlations. In Sec. \ref{Sec:QCMs} we aim to find such a faithful representation by re-interpreting the DAG as a quantum network. We thereby derive a new graph separation rule that does not obey the version of ``Reichenbach's Common Cause Principle", which holds in the classical case, but instead obeys a weaker property we call the ``Quantum Causality Condition". We show that the d-separation can be recovered in a suitably defined classical limit, and we observe that super-quantum correlations (i.e. that exceed Tsirelson's bound) still cannot be explained by our model without fine-tuning. We conclude in Sec. \ref{Sec:Conc} with a discussion about the physical interpretation of the result and possible directions for future work.

\section{Review of classical causal models  \label{Sec:CCM}}

In this section, we review the basic definition of a causal model, here referred to as a \tit{classical causal model} (CCM) to emphasise that it is tied to physical assumptions motivated by classical systems. For more details on causal models and inference, see the book by Pearl and references found therein \cite{PEARL1}.

Before discussing the formal elements of these models, let us briefly recap their historical motivation. In science and statistics, one is often faced with the task of determining the causal relationships between random variables, given some sample data. We might observe that two variables are correlated, but this fact alone does not indicate the direction of the causal influence. If we are limited in our resources, we would like to know which set of follow-up experiments will most efficiently identify the direction of the causal influences, and which causal information can already be deduced from the existing data. Correlations between variables can be represented graphically, for example, we can require that $X$ be independent of $Y$ conditional on a set $Z$ whenever the removal of the nodes $Z$ and their connections from the graph renders the sets of nodes $X$ and $Y$ disconnected in the resulting graph. Such a rule for obtaining independence relations from a graph is referred to as a `graph separation rule'. Such graphs are called \tit{semi-graphoids} and the independence relations they represent satisfy certain axioms, described in Sec \ref{Sec:CCMformal}. 

Correlations can be regarded as restrictions on the possible causal relationships between the variables. Two variables not connected by an edge cannot be directly causally connected, that is, if there is a causal influence of one on the other, it must be mediated by a third set of variables. One can think of causal relations as defining how the observed statistics change after an intervention on a system. When an external agent intervenes to change the probability distribution of some of the variables at will, the distributions of the remaining variables will be updated depending on the direction of the causal influences between them and the manipulated variables; flicking a switch can cause a light to turn off, but extinguishing the light by other means will not affect the position of the switch. Causal information tells us more about the statistical relationships between the variables than can be obtained from correlations alone. It is therefore useful to design a graphical representation and a graph separation rule that captures causal information, not just correlations.

\begin{figure}[!htbp]
\includegraphics[width=8cm]{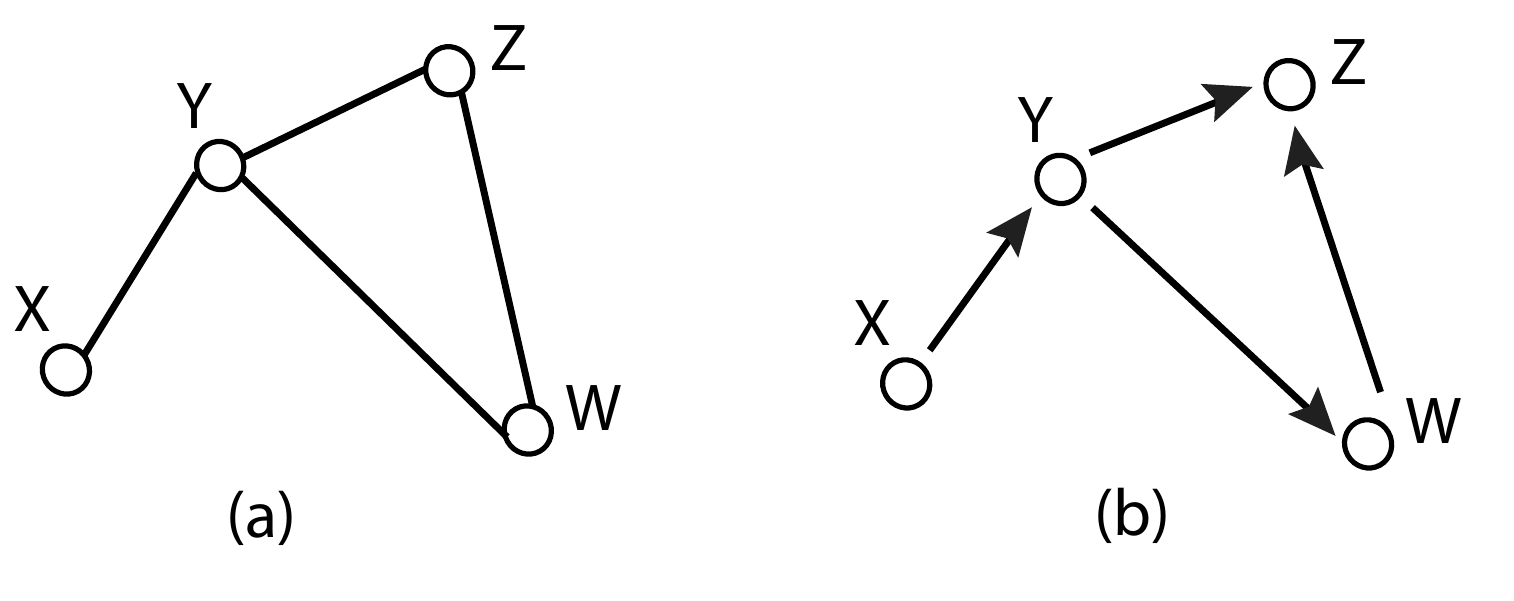}
\caption{(a) A semi-graphoid representing correlations between variables. (b) One possible causal graph consistent with these correlations.}
\label{fig:graphoid}
\end{figure} 
 
The directions of causal influences are represented by adding arrows to the edges in the graph. This supplements the information about correlations with further constraints on the conditional independencies. Every causal graph, up to an absolute ordering of the variables, is in one-to-one correspondence with a list of conditional independence relations called a \tit{causal input list}. The list can be thought of as a set of instructions for generating a probability distribution: one begins by generating the values of the independent variables, then computes the values of any variables that depend directly on them, then the variables that depend on those, and so forth. Hence every causal graph can be taken to represent a stochastic physical process proceeding over many time steps. In practice, working with the causal input list can be cumbersome, so it is more efficient to obtain the conditional independencies directly from the graph using a graph separation rule called \tit{d-separation}. In this work we will propose to upgrade the definitions of causal input list and d-separation to quantum systems.

\subsection{Notation}
Random variables, or sets of random variables, are denoted by capital roman letters, eg. $X,Y, Z$, which take values from a set of possible outcomes. If $\mathcal{E}_X$ is the space of all possible outcomes of $X$, let $P(X)$ represent a probability distribution on $\mathcal{E}_X$ and $P(X=x)$ the probability that the variable $X$ takes the value $x \in \mathcal{E}_X$. In many cases, we will use the term $P(X)$ also to represent $P(X=x)$, except where it might cause confusion. The statement $X=P(X)$ means that the random variable $X$ is distributed over its outcomes according to the distribution $P(X)$. The joint probability $P(X,Y,Z)$ represents a probability distribution over all the possible values of the random variables $\{X,Y,Z\}$. The \tit{conditional} probability $P(X|Y)$ is a set of probability distributions defined on $\mathcal{E}_X$, for the possible values of $Y$. Given a joint distribution $P(X,Y)$, a \tit{marginal} probability $P(Y)$ is defined by summing over all possible values of the other variables, i.e. 
\eqn{
P(Y) &=& \sum_X P(X,Y) \nonumber \\ 
&:=&  \sum_{x \in \mathcal{E}_X} P(X=x,Y) \, .
}
These concepts are united by the law of total probability, which states that $P(X,Y)=P(X|Y) P(Y)$. Unless otherwise specified, we consider only variables with discrete outcome spaces.

\subsection{Formal definitions for causal models \label{Sec:CCMformal} }

Let us consider a set of random variables whose values are governed by some joint probability function and which in general may be correlated. Formally, the statistical dependencies between variables are given by their conditional independence relations:\\

\tbf{Definition \defin{def:CI} \ref{def:CI}: Conditional Independence (CI) relations.} Let $X, Y, Z$ be three disjoint sets of variables. The sets $X$ and $Y$ are said to be \tit{conditionally independent} given $Z$ if knowing $Y$ provides no new information about $X$ given that we already know $Z$ (i.e. $Z$ `screens-off' $Y$ and $X$ from each other). We write this as $(X \LL Y |Z)$, which stands for the assertion that $P(X|Y,Z)=P(X|Z)$. We will often use the shorthand $(X \cup W \LL Z | Y):= (XW \LL Z | Y)$ when dealing with set unions in CI relations. \\

Any joint probability distribution $P$ can be conveniently characterised by the complete set of CI relations that it implies for the variables. In fact, one only needs to specify a subset of CI relations, from which the rest can be obtained using the \tit{semi-graphoid axioms}: \\

 \tbf{Semi-graphoid axioms:}   \\
 \tbf{1.a.} Symmetry: $(X \LL Y |Z) \Leftrightarrow (Y \LL X |Z)$  \\
 \tbf{1.b.} Decomposition:  $(X \LL YW |Z) \Rightarrow (X \LL Y |Z)$   \\
 \tbf{1.c.} Weak union:  $(X \LL YW |Z) \Rightarrow (X \LL Y |ZW)$   \\
 \tbf{1.d.} Contraction: $(X \LL Y |ZW) \& (X \LL W |Z) \Rightarrow (X \LL YW |Z)$  .\\

Note that if $(X \LL Y |Z)$ and $(W \LL Y |Z)$ both hold for disjoint sets $X,Y,Z,W$, then $(XW \LL Y |Z)$ does not necessarily hold. This might seem counter-intuitive, but examples where it fails are easy to construct\footnote{Suppose that all sets represent binary variables $\in \{0,1\}$, and $Y=X\oplus W$ where $\oplus$ is addition modulo 2. Clearly, knowledge of $Y$ does not tell us anything about $X$ or $W$ individually. But knowing $Y$ does reduce the set of possibilities for the joint set $X \cup W$, for example, if $Y=1$ then $X \cup W$ cannot have $X$ and $W$ the same.}. 

The semi-graphoid axioms can be derived directly from the axioms of probability theory. The interpretation of the axioms is  given by the following excerpt from Pearl \cite{PEARL1}, (Chapter 1.1):\\

``The \tit{symmetry} axiom states that, in any state of knowledge $Z$, if $Y$ tells us nothing new about $X$ then $X$ tells us nothing new about $Y$. The \tit{decomposition} axiom asserts that if two combined items of information are judged irrelevant to $X$, then each separate item is irrelevant as well. The \tit{weak union} axiom states that learning irrelevant information $W$ cannot help the irrelevant information $Y$ become relevant to $X$. The \tit{contraction} axiom states that if we judge $W$ irrelevant to $X$ after learning some irrelevant information $Y$, then $W$ must have been irrelevant before we learned $Y$." \\

\tbf{Definition \defin{def:closure} \ref{def:closure}: Semi-graphoid closure.} Given any set $S$ of CI relations, the \tit{closure} of $S$ is the set $\bar{S}$ that includes all CI relations derivable from $S$ using the axioms 1.a-d. \\

Given a joint probability distribution $P$, let $\bar{C}(P)$ denote the complete closed set of CI relations obtainable from $P$. In general, the CI relations do not uniquely fix the probability distribution; there may exist two distinct joint distributions $P$ and $P'$ for which $\bar{C}(P)=\bar{C}(P')$. Hence, the CI relations alone do not capture the full information about the statistics. In the following, we will supplement the CI relations with a causal structure and functional relations in the form of a classical causal model (CCM).

A CCM provides us with an algorithm to generate the statistics of the observables. It can therefore be regarded as an abstract description of a physical system: if the predictions match the actual observations, then the CCM provides a possible explanation of the data. Formally, a CCM consists of two ingredients: an ordered set of CI relations $L_{\mathcal{O}}$, and a set of functions $F$ called the \tit{model parameters}. \\

\tbf{Definition \defin{def:inlist} \ref{def:inlist}: Causal input list.} An ordering $\mathcal{O}$ assigns a unique integer in $\{1,2,...,N \}$ to each member of a set of $N$ variables. Consider an ordered set of variables $\{X_i ; i=1,2,...,N \}$, where the subset of variables $X_j$ with $j<i$ are called the \tit{predecessors} of $X_i$. A \tit{causal input list} is the ordered set of CI relations of the form: $L_{\mathcal{O}} := \{ (X_i \LL R(X_i)| \trm{\tbf{pa}}(X_i) ) : i=1,2,...,N \}$, where each set $\trm{\tbf{pa}}(X_i)$ is a subset of the predecessors of $X_i$ called the \tit{parents} of $X_i$, and $R(X_i)$ are the remaining predecessors of $X_i$ excluding the parents. \\

\tbf{Definition \defin{def:ancdesc} \ref{def:ancdesc}: Ancestors and descendants.} Given a causal input list $L_{\mathcal{O}}$, consider the set of parents of $X_i$, their parents' parents, and so on. These are called the \tit{ancestors} of $X_i$. Similarly, the \tit{descendants} of $X_i$ are all variables for which $X_i$ is an ancestor. We will use $\trm{\tbf{an}}(X)$ to denote the union of the ancestors of a set $X$.\\

\tbf{Definition \defin{def:modparam} \ref{def:modparam}: Model parameters.} Given a causal input list $L_{\mathcal{O}}$, the \tit{model parameters} are a set $F:=\{F_i : i=1,2,...,N \}$ consisting of $N$ probabilistic functions $F_i$. Each $F_i(X)$ is equivalent to applying a deterministic function $f_i(X,U_i)$ with probability $P(U_i)$ for some auxiliary variable $U_i$. The $U_i$ are sometimes called \tit{error variables} and by definition have no parents. Each function $F_i$ determines the probability of $X_i$ conditional on the values of its parents:

\eqn{
\label{modelparam}
P(X_i | \trm{\tbf{pa}}(X_i) ) &=& F_i( \trm{\tbf{pa}}(X_i)) \, \nonumber \\
& =& f_i( \trm{\tbf{pa}}(X_i), U_i) \nonumber \\
&& \trm{with probability } P(U_i) \, .
}
For variables without any parents, called \tit{exogenous variables}, the function $F_i$ just specifies a probability distribution over the possible values of $X_i$, i.e. $F_i(\emptyset) := P(X_i)$. We assume that all exogenous variables, including any error variables $U_i$, are independently distributed (the Markovian assumption).\\

\tbf{Definition \defin{def:CCM} \ref{def:CCM}: Classical causal model.} A \tit{classical causal model} on $N$ variables is a pair $\{ L_{\mathcal{O}} , F \}$ containing a causal input list $L_{\mathcal{O}}$ and model parameters $F$ defined on those variables. Alternatively, a CCM can be specified by the pair $\{ G_L , F \}$, where $G_L$ is the graph generated by $L_{\mathcal{O}}$ (see Sec. \ref{Sec:dsep}).\\  

Given a CCM, we can construct a joint probability by generating random variables in the order specified by $\mathcal{O}$ and using the functions $F$ to define the probabilities of each variable given its parents. These can then be used to construct a joint distribution from the CCM according to the law of total probability:
\eqn{
P(X_1,X_2,...,X_N) = \prod^{N}_{j=1} P(X_j| X_{j+1},...,X_N) \, . 
}

The joint probability obtained in this way from a CCM $\mathcal{M}$ is said to be \tit{generated} by $\mathcal{M}$ and is denoted $P^{\mathcal{M}}$. It satisfies the following property:\\

\tbf{Causal Markov Condition:} Given that $P^{\mathcal{M}}$ is generated by a CCM $\mathcal{M}$, each variable $X_i$ in $P^{\mathcal{M}}$ is independent of its non-descendants, conditional on its parents. \\

\tit{Note:} Our definition of the Causal Markov Condition follows Pearl (Ref. \cite{PEARL1}), in which it is proven to hold for any Markovian causal model (i.e. a model that is acyclic and whose exogenous variables are all independent). In the present work, a CCM is Markovian by construction, so the Causal Markov Condition holds. In the next section, we will use the Causal Markov Condition to motivate interpreting the parents of a variable as its direct causes.\\

\begin{figure}[!htbp]
\includegraphics[width=8cm]{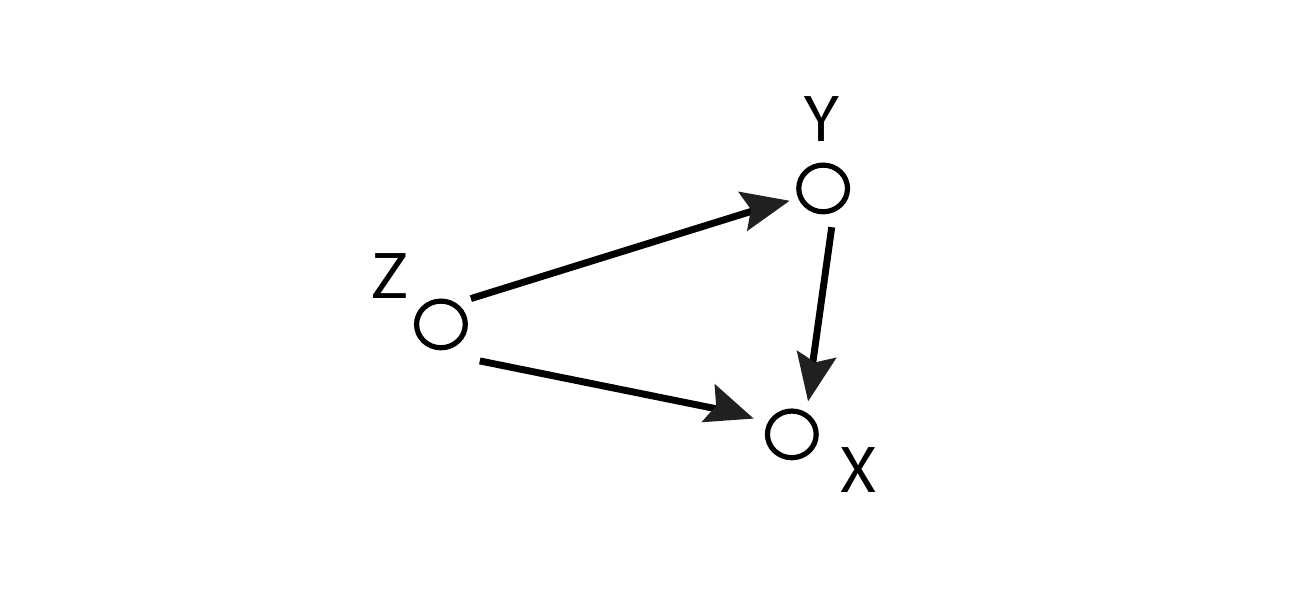}
\caption{A DAG representing a simple classical causal model. $Z$ is a direct cause of both $Y$ and $X$; $Y$ is a direct cause of $X$ only.}
\label{fig:examXYZ}
\end{figure} 

\tbf{Example \exam{exam:xyz} \ref{exam:xyz}: } Consider three variables $X,Y,Z$. Suppose we have the ordering $\mathcal{O}: \{ X,Y,Z \} \rightarrow \{3,2,1\}$, and the causal input list indicates that $\trm{\tbf{pa}}(X)= \{Y, Z\}$; $\trm{\tbf{pa}}(Y)= \{Z\}$; and $\trm{\tbf{pa}}(Z)=\emptyset$. It will be shown in Sec. \ref{Sec:dsep} that this generates the graph shown in Fig. \ref{fig:examXYZ}. Suppose the model parameters are $F=\{f_x,F_y,f_z \}$, where $f_x,f_z$ are deterministic and $F_y(A):=f_y(A,U_Y)$ with probability $P(U_Y)$. Then we obtain the joint probability $P(X,Y,Z)$ as follows: first, generate the lowest variable in the ordering, $Z$, using the random function $P(Z)$. Next, generate $U_Y$ using $P(U_Y)$ and then apply $f_y(Z,U_Y)$ to obtain the value of $Y$. Finally, use $f_x(Y,Z)$ to obtain the value of $X$, the last variable in the ordering. The statistics generated by this procedure are given by:
\eqn{ \label{exampleXYZ}
P(X,Y,Z) &=& \sum_{U_Y} \, P(X|Y,Z) P(Y|Z,U_Y) P(Z) P(U_Y)  \, , \nonumber \\
 \trm{where} &&  \nonumber  \\
P(X|Y,Z) &=& P(X=x |Y=y, Z=z) \nonumber \\
&=& \delta(x, f_x(y,z)) \nonumber \, , \\
 P(Y|Z,U_Y) &=& P(Y=y  | Z=z, U_Y=u_y)   \nonumber \\
&=& \delta(y, f_y(z,u_y)) \nonumber \, . \\
}

\subsection{Physical interpretation \label{Sec:CCMphysical} }

In the previous section, we gave a formal definition of a classical causal model and described how it generates a probability distribution over the outcomes of its random variables. Since these variables represent physical quantities, we would like to supplement this mathematical structure with a physical interpretation of a CCM, as describing the causal relationships between these physical quantities. To do so, we make the following assumption that connects the intuitive concept of a `direct cause' with its mathematical representation.  \\

\tbf{Assumption \assum{ass:directcaus} \ref{ass:directcaus}.} \tit{A variable's parents represent its direct causes.} \\
Physically, we expect that knowledge of the direct causes renders information about indirect causes redundant. Hence the direct causes should \tit{screen off} the indirect causes in the sense of Definition \ref{def:CI}. We therefore define the \tit{direct causes} of $X_i$ as the parents of $X_i$ and the \tit{indirect causes} as the remaining (non-parental) ancestors of $X_i$; the screening-off property then follows from the Causal Markov Condition. \\

The above assumption leads to the following physically intuitive properties of a CCM:  \\

\tbf{Conditioning on common effects:} In a CCM, two variables that are initially independent (i.e. conditional on the empty set) may become dependant conditional on the value of a common descendant. This reflects our intuition that two independent quantities may nevertheless be correlated if one `post-selects' on a future outcome that depends on these quantities. For example, conditional on the fact that two independent coin tosses happened to give the same result, knowing the outcome of one coin toss allows us to deduce the outcome of the other.\\ 

\tbf{Reichenbach's common cause principle (RCCP):} If two variables are initially correlated (i.e. conditional on the empty set) and \tit{causally separated} (neither variable is an ancestor of the other), then they are independent conditional on the set of their \tit{common causes} (parents shared by both variables). \\

It is not immediately obvious that the RCCP follows from the Causal Markov Condition. For a proof using the DAG representation (discussed in the next section) see Ref. \cite{ARNT}. We note that there exist in the literature numerous definitions of the RCCP, so our chosen definition deserves clarification. It was pointed out in Ref. \cite{CAV13} that a general formulation of the principle encompasses two main assumptions. The first states that causally separated correlated variables \tit{must} share a common cause (called the `principle of common cause', or PCC), and the second states that the variables must be screened-off from each other by their common causes (the `factorisation principle' or FP). Our definition of the RCCP refers only to the factorisation property FP, while the PCC is a consequence of the definition of a CCM -- it follows directly from what we have called the assumption of Markovianity and the fact that the variables are only functionally dependant on their parents. By contrast, Ref. \cite{FRITZ} takes the RCCP as being equivalent to the PCC, while the FP is regarded as a separate property that happens to hold for classical correlations. Note that it is precisely the factorisation property that is violated by quantum mechanics, \tit{not} the common cause principle (without conditioning on effects, two independent quantum systems can only become correlated through interaction); so it is not surprising that our framework calls for a rejection of the RCCP, while the definition of Ref. \cite{FRITZ} does not. If one accounts for the difference in definitions, the present work is entirely consistent with Ref. \cite{FRITZ}. We return to this point in Sec. \ref{Sec:QCMs}. \\

Finally, we note that one can interpret the ordering of variables $\mathcal{O}$ as representing the time-ordering of the variables, such that each variable represents a physical quantity localised to an event in space-time. However, this interpretation is not strictly necessary for what follows. Indeed, it may be interesting to consider alternative interpretations in which some causal influences run counter to the direction of physical time, such as in the retro-causal interpretation of quantum mechanics \cite{PRI08}. 

\subsection{The DAG representation of a CCM \label{Sec:dsep}}

It is useful to represent $L_{\mathcal{O}}$ using a \tit{directed acyclic graph} (DAG), which can be thought of as the causal `skeleton' of the model. In the DAG representation of $L_{\mathcal{O}}$, there is a node representing each variable and a directed arrow pointing to the node from each of its parents. The DAG $G_L$ constructed in this way is said to be \tit{generated} by the causal input list $L_{\mathcal{O}}$. The parents of a node in a DAG are precisely those nodes that are directly connected to it by arrows pointing towards it. It is straightforward to see that the ancestors of $X$ are represented by nodes in the DAG that each have a directed path leading to $X$, and the descendants of $X$ are those nodes that can be reached by a directed path from $X$. In Example \ref{exam:xyz}, the system is represented by the DAG shown in Fig. \ref{fig:examXYZ}. 

\begin{figure}[!htbp]
\includegraphics[width=8cm]{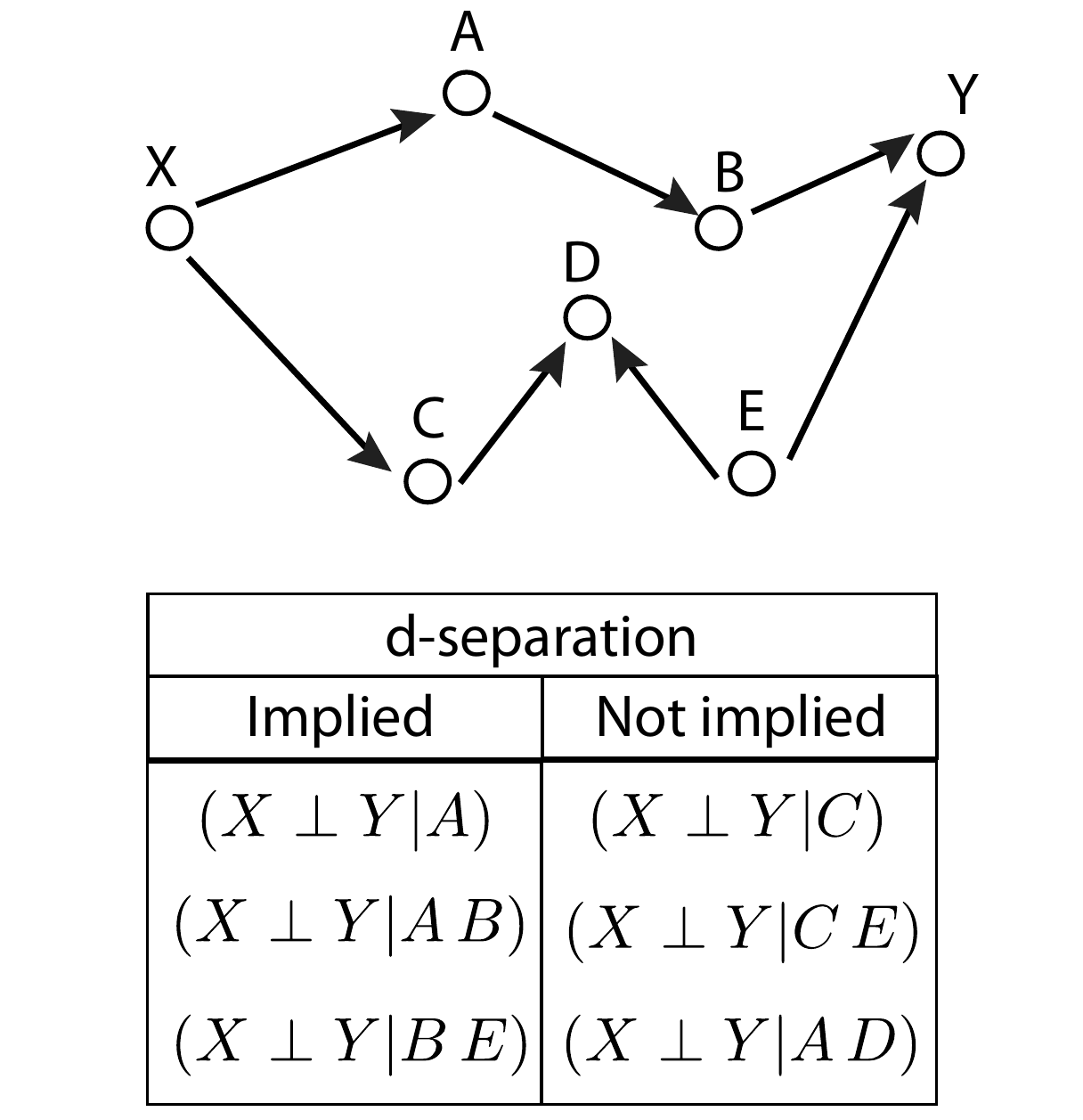}
\caption{An illustration of d-separation. The table indicates which CI relations are implied by d-separation, and which are not.}
\label{fig:dsep}
\end{figure}

To establish a correspondence between a DAG and its generating causal input list $L_{\mathcal{O}}$, we need an algorithm for reconstructing a list of CI relations $L_{\mathcal{O}}$ from a DAG, such that the list generates the DAG. For this purpose, one uses \tit{d-separation} (see Fig. \ref{fig:dsep}).\\

\tbf{Definition \defin{def:dsep} \ref{def:dsep}: d-separation.} Given a set of variables connected in a DAG, two disjoint sets of variables $X$ and $Y$ are said to be \tit{d-separated} by a third disjoint set $Z$, denoted $(X \LL Y | Z)_d$, if and only if every undirected path (i.e. a path connecting two nodes through the DAG, ignoring the direction of arrows) connecting a member of $X$ to a member of $Y$ is \tit{rendered inactive} by a member of $Z$. A path connecting two nodes is rendered inactive by a member of $Z$ if and only if:\\
(i) the path contains a chain $i \rightarrow m \rightarrow j$ or a fork $i \leftarrow m \rightarrow j$ such that the middle node $m$ is in $Z$, or\\
(ii) the path contains an inverted fork (\tit{head-to-head}) $i \rightarrow m \leftarrow j$ such that the node $m$ is not in $Z$, and there is no directed path from $m$ to any member of $Z$.  \\
A path that is not rendered inactive by $Z$ is said to be \tit{active}.\\

By assuming that all d-separated nodes are independent conditional on the separating set (see below), we can then obtain CI relations from the DAG. In principle, the rules for d-separation can be derived from the requirement that it produces the CI relations contained in the list $L_{\mathcal{O}}$ that generates the DAG. However, d-separation also provides an intuitive graphical representation of the physical principles discussed in the previous section. In particular, a path between two nodes in the graph is rendered inactive by a set $Z$ in precisely those situations where we would physically expect the two variables to be independent conditional on $Z$: when we are not conditioning on any common effects (head-to-head nodes); when we are conditioning on a common cause (as in the RCCP); or when we are conditioning on a node that is a link in a causal chain (screening off indirect causes). With the physical interpretation in mind, we are motivated to use d-separation to obtain CI relations using the correspondence:
\eqn{
(X \LL Y| Z)_d \Rightarrow (X \LL Y | Z) \, ,
}
i.e. we assume that if $X$ and $Y$ are d-separated by $Z$ in a DAG, then they are conditionally independent given $Z$ in the semi-graphoid closure of any list $L_{\mathcal{O}}$ that generates the DAG. 

Formally, let $G$ be a DAG, and let $C(G)$ be the set of CI relations obtainable from $G$ using d-separation, and $\bar{C}(G)$ the closure of this set. We then have the following theorem:\\

\tbf{Theorem \thrm{thrm:dsep} \ref{thrm:dsep} (Verma \& Pearl \cite{VERMA}):}\\
Let $G_L$ be the DAG generated by $L_{\mathcal{O}}$. Then $\bar{C}(G_L) = \bar{L}_{\mathcal{O}}$. That is, the closure of the DAG is equal to the closure of the causal input list, so that every CI relation implied by the DAG $G_L$ also follows logically from $L_{\mathcal{O}}$ and vice-versa.\\ 

Theorem \ref{thrm:dsep} implies that d-separation is \tit{sound}, since every CI relation obtainable from the DAG is in the closure of the causal input list ($\bar{C}(G_L) \subseteq \bar{L}_{\mathcal{O}}$), and \tit{complete}, since there are no CI relations implied by the causal input list that are not obtainable from the DAG ($\bar{L}_{\mathcal{O}} \subseteq \bar{C}(G_L) $).\\

Given a DAG $G$, one can always find a causal input list that generates $G$ as follows: choose a total ordering $\mathcal{O}$ that is consistent with the partial ordering imposed by $G$, and then write down the ordered list of CI relations of the form $(X_i \LL R(X_i)|\trm{\tbf{pa}}(X_i))$, where the parents of each variable are the same as the parents of its representative node in $G$. Moreover, the list obtained in this way is unique, modulo some freedom in the ordering of causally separated events (eg. if the variables represent events in relativistic spacetime, this freedom corresponds to a choice of reference frame). It is this feature that allows us to replace the causal input list with its corresponding DAG in the definition of a CCM (Definition \ref{def:CCM}).

\begin{figure}[!htbp]
\includegraphics[width=8cm]{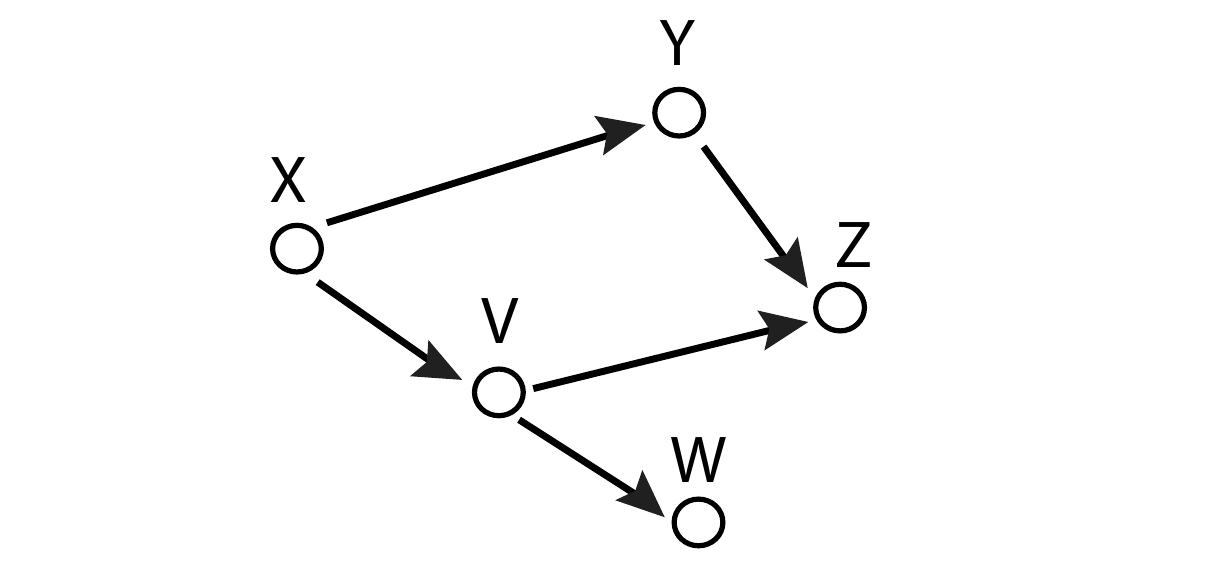}
\caption{A DAG can also be used to construct a causal input list that generates it (see text).}
\label{fig:daglo}
\end{figure} 

\tbf{Example \exam{exam:daglo} \ref{exam:daglo}: } Consider the DAG shown in Fig. \ref{fig:daglo}. The graph assigns parents as follows: 
\eqn{
\trm{\tbf{pa}}(X) &=& \emptyset \, , \nonumber \\
  \trm{\tbf{pa}}(V) &=& X \, , \nonumber \\
  \trm{\tbf{pa}}(Y) &=& X  \, , \nonumber \\
   \trm{\tbf{pa}}(W) &=& V  \, , \nonumber \\ 
    \trm{\tbf{pa}}(Z) &=& V \cup Y  \,. 
}

A total ordering consistent with this graph is $\mathcal{O}: \{ X,V,W,Y,Z \} \rightarrow \{1,2,3,4,5\}$. Therefore, we obtain the causal input list:

\eqn{
L_{\mathcal{O}} &=& \{(X \LL \emptyset | \emptyset), (V \LL \emptyset | X), (W \LL X| V), \nonumber \\
&& (Y \LL \emptyset | X), (Z \LL X W | Y V) \} \, .
}  

Finally, we have the following useful definitions:\\

\tbf{Definition \defin{def:indep} \ref{def:indep}: Independence maps and perfect maps.}
Given $P(X)$ and a DAG $G$ on the same variables, we will call $G$ an \tit{independence map} (I-map) of $P$ iff $\bar{C}(G) \subseteq \bar{C}(P)$. If equality holds, $\bar{C}(G) = \bar{C}(P)$, then $G$ is called a \tit{perfect map} of $P$. \\

The intuition behind this definition can be understood as follows. A DAG is an independence map of a probability distribution iff every CI relation implied by the DAG is also satisfied by the distribution. That means that if two variables are not causally linked in the DAG, they must be conditionally independent in the distribution. However, the converse need not hold: the arrows in a DAG represent only the \tit{possibility} of a causal influence. In general, depending on the choice of model parameters, it is possible for two variables to be connected by an arrow and yet still be conditionally independent in the probability distribution. Equivalently, one can find a probability distribution that satisfies \tit{more} conditional independencies than those implied by its DAG. A DAG is a perfect map iff it captures \tit{all} of the CI relations in the given distribution, i.e. every causal dependence implied by the arrows in the DAG is manifest as an observed signal in the statistics.

Interestingly, there exist distributions for which no DAG is a perfect map, the key example being any bipartite distribution that violates a Bell inequality. This fact forms the basis for the criterion of \tit{faithfulness} of a CCM, discussed in the following section.

\subsection{Faithful explanations and fine-tuning \label{SecExpl}}

Suppose we obtain the values of some physical observables over many runs of an experiment and that the statistics can be modelled (to within experimental errors) by a CCM. Then we can say that the CCM provides a causal explanation of the data, in the sense that it tells us how physical signals and information propagate between the physical observables. In particular, it allows us to answer counterfactual questions (what would have happened if this observable had taken a different value?) and predict how the system will respond to interventions (how will other quantities be affected if a given variable is forcibly altered?). These notions can be given a rigorous meaning using causal models, and they constitute a formal framework for making causal inferences from observed data. In the present work, we will be primarily concerned with defining a quantum causal model that can be given a useful graphical representation in DAGs, so we will not discuss interventions and inference in causal models (the interested reader is referred to Ref. \cite{PEARL1} for inference in the classical case). 

Before we consider quantum systems, it will be useful to review some caveats to the question of whether a CCM provides an adequate description of some given data, and in particular, whether a given CCM gives a \tit{faithful} account of the observed statistics. The first caveat has to do with the possibility of hidden, or \tit{latent} variables. Suppose that we have a probability distribution $P(X)$ for which there is no CCM that generates it (this can occur, for example, if some exogenous variables in the model are found to be correlated with each other, thereby violating the basic property of Markovianity required for a CCM). Rather than giving up the possibility of a causal explanation, we might consider that there exist additional variables that have not been observed, but whose inclusion in the data would render the statistics explainable by some CCM. Formally, suppose there is an extension of $P(X)$ to some larger distribution $P'(X,\lambda)$ that includes latent variables $\lambda$, such that the observed statistics are the marginal probabilities obtained by summing over the unobserved variables:

\eqn{P(X)=\sum_{\lambda} P'(X,\lambda)}

If there exists a CCM $\mathcal{M}(X,\lambda)$ such that $P^{\mathcal{M}}(X,\lambda)=P'(X,\lambda)$, then we can say that this CCM explains the probability distribution $P(X)$ with the aid of the latent variables $\lambda$. The admittance of hidden variables in causal models seems to lead to a problem: it turns out that \tit{every} probability distribution $P(X)$ can be explained by a CCM, with the aid of a sufficient number of hidden variables! For this reason, we further constrain the possible explanations of the observed data by requiring that the models be \tit{faithful} to the data:\\

\tbf{Definition \defin{def:faith} \ref{def:faith}: Faithfulness.} Consider a distribution $P(X)$ and a CCM $\mathcal{M}(X)=\{G, F \}$ that generates $P(X)$. The explanation offered by $\mathcal{M}(X)$ is called \tit{faithful} to $P(X)$ iff the DAG derived from $\mathcal{M}(X)$ is a perfect map of $P(X)$, i.e. $\bar{C}(G) = \bar{C}(P)$. \\ 
\tit{Latent variables:} Suppose there is no CCM $\mathcal{M}(X)$ that is faithful to $P(X)$. Consider instead a CCM $\mathcal{M}(X,\lambda)=\{G' , F' \}$, which obtains $P(X)$ by summing the generated distribution $P'(X,\lambda)$ over the hidden variables $\lambda$. This extended CCM is considered faithful to $P(X)$ iff every CI relation in $P(X)$ is implied by the extended DAG $G'$, i.e. $\bar{C}(P) \subseteq \bar{C}(G')$.\\

The motivation for this definition is that a faithful explanation of the observed statistics is a better candidate for describing the `real causal structure' of the system than an unfaithful explanation, because it accurately captures all causal dependencies in the observed statistics. If there exists no faithful explanation of the observed statistics, but one can obtain a faithful explanation using hidden variables, then we can interpret the statistics of the observed variables as the marginal statistics arising from ignoring the unobserved variables. Note that not all probability distributions can be \tit{faithfully} reproduced by some CCM, even with the aid of hidden variables.  \\

Geiger \& Pearl \cite{GEIGER} showed that, for every DAG $G$, one can explicitly construct a distribution $P$ such that $G$ is faithful for $P$, i.e. such that $\bar{C}(G) = \bar{C}(P)$ holds. Furthermore, it can be shown that if there exists a DAG $G$ that is faithful for a given $P$, then there must exist a set of model parameters $F$ such that the CCM $\{ G, F \}$ generates $P$ \cite{DRU}. However, we have already mentioned that there exist probability distributions for which no DAG is a perfect map; this will be relevant when we consider quantum mechanics in Sec. \ref{Sec:QMintro}.  

Finally, faithfulness can be equivalently defined as the rejection of \tit{fine-tuning} of the model parameters, as noted in Refs. \cite{PEARL1,WOOD}. In particular, if a DAG is unfaithful, this implies that there exists at least one conditional independence in the statistics that is not implied by the DAG. It can be proven that within the set of probability distributions compatible with the DAG, those that satisfy such additional CI relations are a set of measure zero \cite{PEARL1}. Thus, the only way that such additional CI relations could arise is by a kind of `conspiracy' of the model parameters $F$ to ensure that the extra conditional independence holds, even though it is not indicated by the causal structure. In this sense, fine-tuning represents causal influences that exist at the ontological level but cannot be used for signalling at the level of observed statistics, due to the careful selection of the functional parameters; any small perturbation to these parameters would result in a signal.  \\

\tbf{Example \exam{exam:finet} \ref{exam:finet}: } Recall the system of Example \ref{exam:xyz} and suppose that we observe $(X \LL Y|Z)$ in $P$. This CI relation is not found in $\bar{L}_{\mathcal{O}}$; in fact, there is a directed edge from $Y$ to $X$ in the DAG of $L_{\mathcal{O}}$ (Fig. \ref{fig:examXYZ}). The only way to account for this discrepancy is if the model parameters $F$ are chosen such that the predicted signal from $Y$ to $X$ is obscured. This could happen if $X,Y,Z$ are positive integers and we choose model parameters $f_y(z,u_y)=u_y+z$ and $f_x(y,z)=y+z-k$ for some integer constant $k$. For simplicity, suppose the model is deterministic, $P(U_Y=u_y)=1$. If the numbers $u_y$ and $k$ just so happen to be equal, then we obtain the joint probability:
\eqn{
P(X,Y,Z)&=& \delta(x, f_x(\, f_y(z,k) \, , z) \, ) P(Z) \nonumber \, , \\
&=& \delta(x, 2z) P(Z) \, ,
}
and the value of $X$ tells us nothing about the value of $Y$, because the $k$'s conveniently cancel out, leading to the observed independence $(X \LL Y|Z)$. This can be understood as a coincidence in the model parameters $f_x$ and $f_y$, whereby they each depend on the same parameter $k$. Indeed, if the constants $k$ and $u_y$ were allowed to differ in each function, then the cancellation would not occur, and $X$ would still carry information about $Y$ in accordance with the causal structure. Hence, absence of fine-tuning in $P$ with respect to a causal input list $L_{\mathcal{O}}$ can also be defined as the requirement that the CI relations observed in $P$ should be robust under changes in the model parameters consistent with $L_{\mathcal{O}}$ \cite{PEARL1}. 

\subsection{Does quantum mechanics require fine-tuning? \label{Sec:QMintro}}

Consider a probability distribution $P(A,B,S,T)$, satisfying the generating set of CI relations $K:=\{(S \LL T| \emptyset),\,(A \LL T| S),\,(B \LL S| T) \}$. Suppose that the closure of this set contains all the CI relations satisfied by $P$, i.e. $\bar{K}=\bar{C}(P)$. This represents a generic Bell-type experiment: the setting variables $S,T$ are independent of each other, and there is no signalling from $S$ to $B$ or from $T$ to $A$, but the outputs $A,B$ are correlated. (Here, the absence of signalling is only a constraint on the allowed probability distributions. It refers to the fact that the marginal distribution of the outcomes on each side must be conditionally independent of the values of the setting variables on the opposite side, a requirement often called `signal locality' in the literature. It does not forbid the possibility of signalling at the ontological level, as in non-local hidden variables, etc.)  

Note that we cannot explain these correlations by a CCM without latent variables, because $A$ and $B$ are correlated without a common cause. Hence let us consider the extended distribution $P'(A,B,S,T,\lambda)$ satisfying $\sum_{\lambda} P'(A,B,S,T,\lambda) = P(A,B,S,T)$. Of course, we require that $\bar{K} \subseteq \bar{C}(P')$, but we can impose additional physical constraints on the hidden variable $\lambda$. In particular, we expect $\lambda$ to be independent of the settings $S,T$ and, in keeping with the no-signalling constraint, to represent a common cause subject to Reichenbach's principle. This leads to the extended set of constraints $K':=\{(S\LL T \lambda | \emptyset),(T \LL S \lambda |\emptyset),\,(A \LL T| S\lambda),\,(B \LL S |T\lambda) \}$ and we assume that $\bar{K'} \subseteq \bar{C}(P')$, i.e. that the extended distribution satisfies at least these constraints. We then ask whether there exists a CCM that can \tit{faithfully} explain the observed correlations. 

\begin{figure}[!htbp]
\includegraphics[width=8cm]{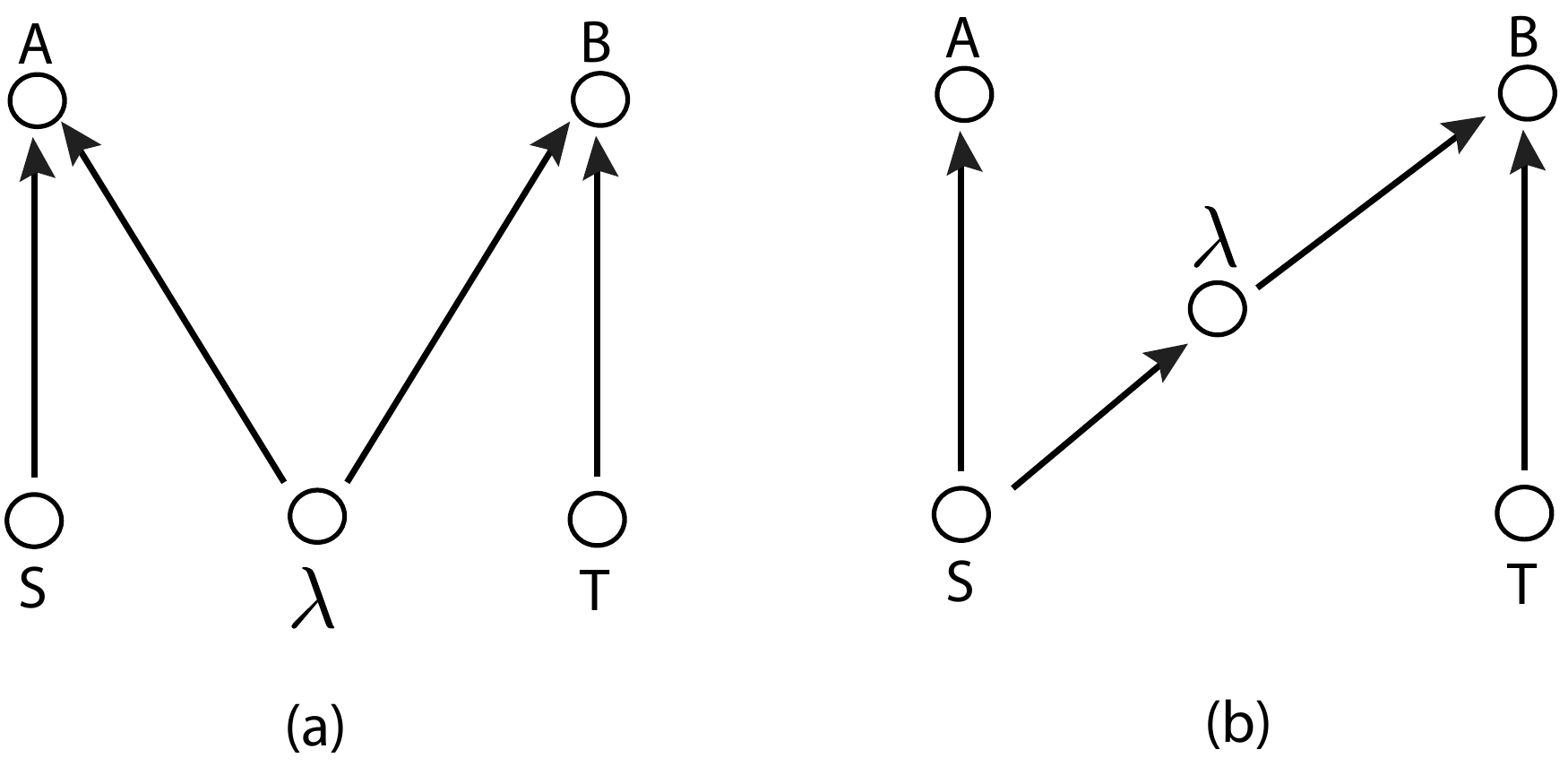}
\caption{(a) A possible DAG for a Bell-type experiment using hidden variables. (b) A DAG that can serve as an unfaithful explanation of Bell-inequality violation.}
\label{fig:Bell}
\end{figure}

If $A$ and $B$ are independent conditional on the hidden variable $\lambda$ in the distribution $P'$, i.e. if $(A \LL B| \lambda)$ holds in $\bar{C}(P')$, then it is easy to see that $\lambda$ qualifies as a common cause of $A$ and $B$ and the correlations can be explained by a CCM with the DAG $G'$ shown in Fig. \ref{fig:Bell} (a). It can be shown that this occurs whenever $P$ satisfies Bell's inequality \cite{WOOD}. Conversely, Wood \& Spekkens showed that if $P$ violates Bell's inequality, there is no CCM that can faithfully explain $P$, even allowing hidden variables. Of course, one can find numerous un-faithful explanations, such as the DAG in Fig. \ref{fig:Bell} (b) and fine-tuning of the model parameters to conceal the causal influence of $S$ on $B$. This result implies that, in general, CCMs cannot faithfully explain the correlations seen in entangled quantum systems.\\

\tbf{Example \exam{exam:bohm} \ref{exam:bohm}: } Consider the deBroglie-Bohm interpretation of quantum mechanics. This interpretation gives a causal account of Bell inequality violation using super-luminal influences, one possible variant of which is depicted in Fig. \ref{fig:Bell} (b). Here, $\lambda$ is a hidden variable that carries information about the setting $S$ faster than light to the outcome $B$. The model posits a CCM that generates a distribution $P'(A,B,S,T,\lambda)$ and the observed statistics $P(A,B,S,T)$ are interpreted as the marginal obtained from this distribution by summing over $\lambda$. The no-signaling CI relations $(A \LL T | S), (B \LL S | T)$ hold in the observed statistics, however they do not follow from the DAG Fig. \ref{fig:Bell} (b) which includes the hidden variables, hence the CCM that generates $P'$ using this graph is not a faithful explanation for $P$. In general, the deBroglie-Bohm interpretation and its variants appear to require fine-tuning \cite{WOOD}.\\

How should we interpret this result? On one hand, we might take it as an indication that faithfulness is too strong a constraint on the laws of physics, and that nature allows hidden variables whose causal influences are concealed at the statistical level by fine-tuning. Alternatively, we could take it to indicate that the class of physical models describable by CCMs is not universal, and that a new type of causal model is needed to give a faithful account of quantum systems. Along these lines, we could choose to interpret Fig. \ref{fig:Bell} (a) as a quantum circuit, where $\lambda$ now stands for the preparation of an entangled pair of quantum systems and the arrows stand for their distribution across space. In doing so, we implicitly shift our perception of quantum mechanics from something that needs to be explained, to something that forms part of the explanatory structure. We no longer seek to explain quantum correlations by an underlying causal mechanism, but instead we incorporate them as a fundamental new addition to our causal structure, which can then be used to model general physical systems. This approach entails that we no longer require $(A \LL B| \lambda)$ to hold for the ``common cause" $\lambda$ and hence that we abandon Reichenbach's Common Cause Principle (specifically, the factorisation property). This in turn implies that d-separation is no longer the correct criterion for reading CI relations from the DAG, when the DAG is interpreted as a quantum circuit. In what follows, we will propose a new criterion that serves this purpose, leading to the concept of a quantum causal model.

\section{Quantum Causal Models \label{Sec:QCMs}}

\subsection{Preliminaries \label{Sec:Qprelims}}

We begin by considering quantum networks modelled as a DAGs, in which the nodes represent state preparations, unitary transformations and measurements. Based on this interpretation, we obtain a corresponding notion of a quantum input list and a graph separation criterion that connects the DAG to the list that generates it. We mention that there exist other approaches to quantum computation in which it would be interesting to explore causal relations, such as measurement-based quantum computation. For efforts along these lines, see eg. \cite{TOM14, MIA13}.

The general theory of quantum networks as given in Ref. \cite{CHI} provides a DAG representation in which nodes represent completely general quantum operations. Below, we define a canonical form of a general quantum network in order to cleanly separate the classical apparatus settings from the measurement outcomes, to facilitate the definition of a graph separation criterion. Given a DAG, we divide the nodes into four classes: as before, those with no incoming edges are called \tit{exogenous}; those with no outgoing edges are called \tit{drains}; those with ingoing and outgoing edges are called \tit{intermediates}. We assign the following interpretations to the elements of the DAG:\\

\tbf{Edges:} every edge in a DAG is associated with a Hilbert space of dimension $2^n$, with $n \geq 1$. Thus, we can associate an integer number of qubits with each edge in the graph. The Hilbert space dimension is allowed to be different for different edges; however, for intermediate nodes we require that the total Hilbert space dimension of the ingoing edges (obtained by multiplying the dimensions of the individual edges) be equal to the total dimension of the outgoing edges. The reason for this constraint is that it allows us to associate a unitary map to each intermediate node.\\

\tbf{Exogenous nodes:} Every exogenous node is associated with a random variable. Each possible value of the variable corresponds to the preparation of a normalised pure state (a source). The set of pure states need not be orthogonal - in fact they may even be degenerate, with more than one value of the variable corresponding to preparation of the same state. The only requirement is that the states exist in a Hilbert space with dimension equal to $\mathcal{H}^{\trm{(out)}}$, which is the tensor product of the Hilbert spaces of all the outgoing edges.\\

\tbf{Drains:} Every drain is associated with a random variable. Each value of the variable corresponds to the outcome of a projective measurement on $\mathcal{H}^{\trm{(in)}}$, the tensor product of the Hilbert spaces of all the ingoing edges. The measurement basis is assumed to be fixed by convention. For example, since the dimension of $\mathcal{H}^{\trm{(in)}}$ is $2^M$ for some integer $M \geq 1$, we can always take the measurement basis to be the computational basis of $M$ qubits. \\

\tbf{Intermediate nodes:} Every node with both incoming and outgoing edges is associated with a random variable. Each value of the variable represents a unitary operator on $\mathcal{H}^{\trm{(in)}} (=\mathcal{H}^{\trm{(out)}})$. \\

The above definitions allow us to associate a quantum network to any DAG. Conversely, every quantum network has a representation as a DAG of this form.\\

\begin{figure}[!htbp]
\includegraphics[width=8cm]{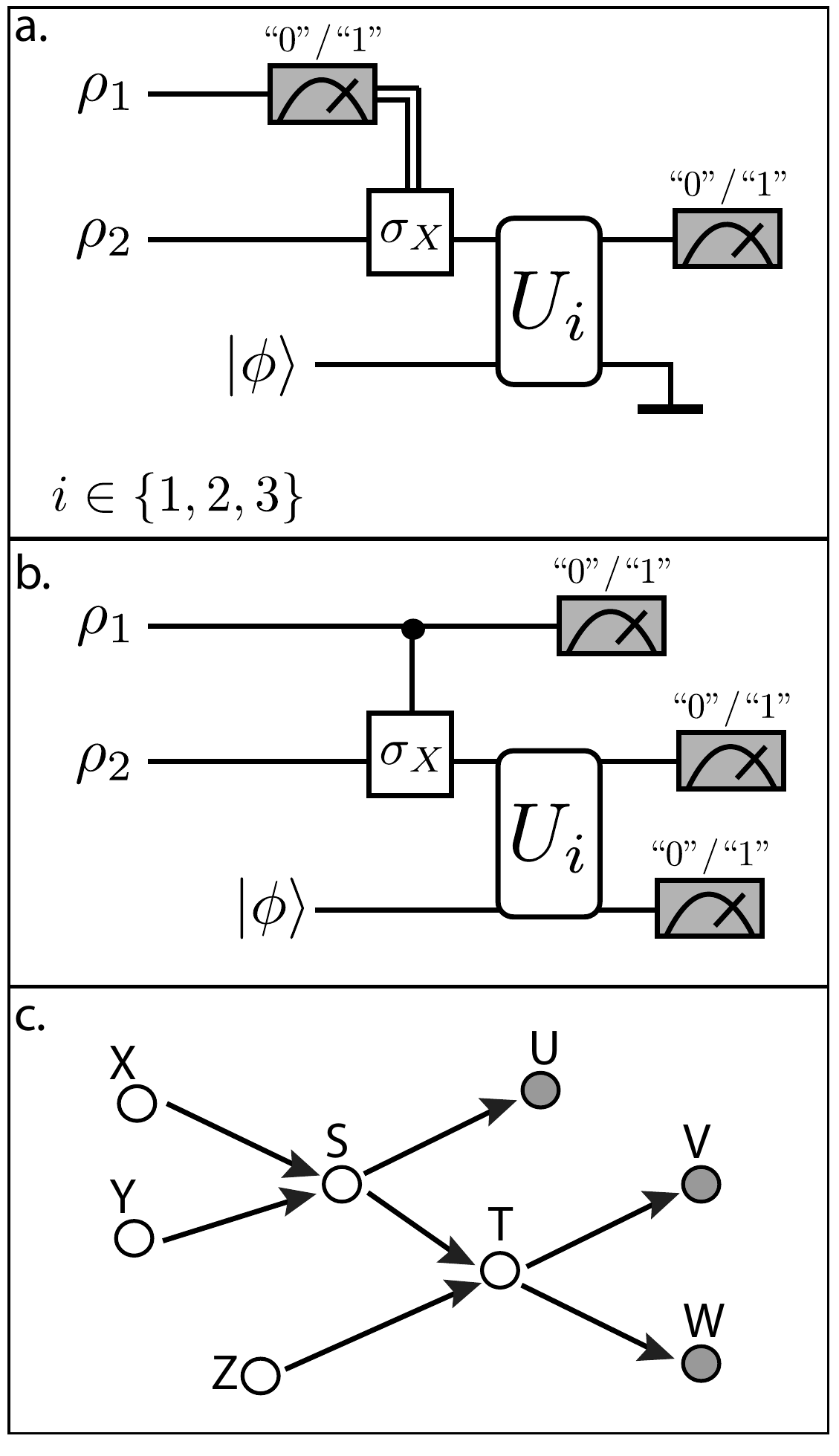}
\caption{The conversion of a quantum circuit (a) into a DAG, using the recipe outlined in the text. The conversion is performed by replacing feed-forwards with unitary interactions, followed by measurements in a fixed basis; and replacing general measurements with unitary coupling to ancilla states, followed by measurement in a fixed basis, as shown in (b). The resulting DAG is shown in (c).}
\label{fig:ffwd}
\end{figure} 

\tbf{Example \exam{exam:dagcirc} \ref{exam:dagcirc}: } Consider the circuit in Fig. \ref{fig:ffwd} (a). This describes the preparation of two qubits as mixtures $\rho_1=\gamma_1 \densop{\psi_0}{\psi_0} + (1-\gamma_1) \densop{\psi_1}{\psi_1}$ and $\rho_2=\gamma_2 \densop{\psi_0}{\psi_0} + (1-\gamma_2) \densop{\psi_1}{\psi_1}$ in an arbitrary orthogonal basis $\{ \ket{\psi_0}, \ket{\psi_1} \}$. These are followed by a measurement of the first qubit in the computational $\sigma_Z$ basis $\{ \ket{0}, \ket{1} \}$ and the subsequent application of a $\sigma_X$ gate to the second qubit, conditional on the outcome of the first measurement. Finally a POVM is applied to the second qubit by coupling it via unitary interaction (either $U_1,U_2$ or $U_3$) to a third ancilla qubit $\ket{\phi}$. The ancilla is traced out and the remaining qubit measured in the $\sigma_Z$ basis. In Fig. \ref{fig:ffwd} (b) the feed-forward has been replaced with a unitary interaction (a CNOT) followed by tracing out the first qubit (all feed-forwards can be described in this way to ensure that the setting variables, representing the choice of input state and unitary, remain independent of each other). The tracing-out of the ancilla qubit is replaced with a measurement in the $\sigma_Z$ basis, whose outcome can be ignored. In this form, the circuit can be cast directly into a DAG, as shown in Fig. \ref{fig:ffwd} (c). The variables $X$ and $Y$ take values corresponding to the basis states $\{ \ket{0}, \ket{1} \}$, distributed with probabilities so as to produce the mixed states $\rho_1,\rho_2$. $Z$ and $S$ are single-valued, corresponding to the state $\ket{\phi}$ and the unitary CNOT respectively. $T$ has three values corresponding to the three possible unitaries, and is distributed according to the probability of each unitary being implemented. Finally, $U,V,W$ are all binary-valued, corresponding to outcomes $\in \{0,1\}$ and whose probabilities are given by quantum mechanics (see Sec \ref{Sec:Qinput}). 

Note that a single DAG can represent \tit{any} member of the class of quantum networks with the same basic topology (i.e. the same connections between preparations, unitaries and measurements). Thus, in a quantum causal model, the preparations and unitaries are taken as the model parameters and the DAG provides the causal structure, as explained in the next section. \\ 

\subsection{Quantum input lists and model parameters \label{Sec:Qinput}}

Recall that the classical causal input list $L_{\mathcal{O}}$ represents a set of conditional independence relations between variables in a CCM, from which a DAG can be easily constructed. The motivation for the causal input list comes from its physical interpretation, discussed in Sec. \ref{Sec:CCMphysical}, which embodies principles like the RCCP that we expect to hold for classical physical systems. Hence, to define the quantum analog of a causal input list, we should begin by asking: \tit{for variables in a quantum network, what physical principles constrain their statistical dependencies?} 

First, we note that the observables in a quantum network fall naturally into two distinct categories: \tit{settings} $S_i$ and \tit{outcomes} $O_i$ (we continue to use $X$ to denote a generic variable or set of variables). The settings $S_i$ determine the states produced at the sources and the unitaries applied in the network, while the $O_i$ represent the outcomes of measurements in a fixed basis. Since the settings $S_i$ play the same role as the exogenous variables in a CCM, we assume that they are all distributed independently of each other; however, unlike in a CCM, this property now also applies to variables represented by intermediate nodes. This assumption is the analog of the Markovianity assumption for a CCM, and it embodies one aspect of the common cause principle that \tit{is} retained in quantum mechanics, namely, that correlated variables (conditional on the empty set) must share a common source, or must have interacted previously. It is in this sense that the RCCP can be said to hold for quantum correlations in Ref. \cite{FRITZ} (recall the discussion of Sec. \ref{Sec:CCMphysical}).

Also as before, we assume an absolute ordering of the variables and enforce the physical assumption of causality (no causal loops) and we again assign a set of parents to each variable, representing the connections in the network and (implicitly) the possibility of a causal influence. However, unlike the case of a CCM, we are not able to interpret the parents of a variable as its direct causes. This is because the values of the settings by definition do not have any causes in the network (they are chosen by external factors, like experimental intervention). Furthermore, the parents of an outcome no longer screen it off from its other ancestors: the influence of an initial state preparation on the measurement outcome cannot in general be screened off by a choice of intermediate unitary. We leave it as an open question whether one can formulate a quantum network in a manner that respects this property of CCMs; we will find it more convenient simply to abandon it. Indeed, since the variables representing the preparation and choice of unitary are assumed to be independent, they cannot carry any information about each other, nor can the variable representing the unitary reveal any information about the quantum system on which it acts. The assignment of parents to the variables therefore places much weaker constraints on the correlations than in the classical case. However, the following physical assumption is still justified in a quantum network:\\

\tbf{Assumption \assum{ass:pastcone} \ref{ass:pastcone}.} \tit{The possible causes of an outcome are its ancestors.} \\

This assumption reflects our intuition that it is only the operations performed on a quantum system leading up to its measurement that can have a causal effect the measurement outcome. Indeed, it is also argued in Ref. \cite{FRITZ} that there is no reason to maintain the distinction between direct and indirect causes in any generalised model that goes beyond classical correlations.\\

\noindent It is clear from our discussions in Sec. \ref{Sec:CCM} that the Causal Markov Condition is not expected to hold in a quantum network, since the RCCP no longer holds. Instead, we expect it to be replaced by a weaker property:\\

\tbf{Quantum Causality Condition:} An outcome is independent (conditional on the empty set) of all settings that are not its causes and all outcomes that do not share a common cause.  \\ 

This property expresses the fact that outcomes should be independent of any settings from which they are causally disconnected and should be correlated only with other outcomes that share a common cause. This property holds also in the classical case, but unlike the classical case, we now do not require sets of outcomes to be independent of each other \tit{conditional} on their common causes -- instead we allow them to still be dependent, admitting violations of the factorisation property of the RCCP (recall Sec. \ref{Sec:CCMphysical}). In addition to this weakened version of the RCCP, we still have the classical feature that independent variables can become dependant conditional on common effects. Thus, for two variables to be independent, we will still have to avoid conditioning on certain colliders. \\  

To make these ideas formal, let us consider a set of random variables partitioned into \tit{outcomes} $O$ and \tit{settings} $S$. The following definitions will also be useful:\\

\tbf{Definition \defin{def:chain} \ref{def:chain}: $O'$-chain.} Given a set of outcomes $O'$, two other sets of outcomes $O_1$ and $O_2$ are said to be \tit{connected by an} $O'$\tit{-chain} iff $O_1$ shares an ancestor with a member of $O'$ that shares an ancestor with another member of $O'$, (etc), that shares an ancestor with $O_2$. A set of settings $S_1$ is linked to $O_2$ by an $O'$-chain iff $S_1$ has a descendant in $O'$ that is connected by an $O'$-chain to $O_2$. Similarly, $S_1$ and $S_2$ are connected by an $O'$-chain iff they both have descendants in $O'$ that are connected in this way. \\
 
\tbf{Definition \defin{def:detached} \ref{def:detached}: $O'$-detached.} Given a set of outcomes $O'$ and some variables $V$, the set of all variables not connected to $V$ by an $O'$-chain are said to be $O'$\tit{-detached} from $V$, denoted $\trm{\tbf{dt}}_{O'}(V)$.\\

\begin{figure}[!htbp]
\includegraphics[width=8cm]{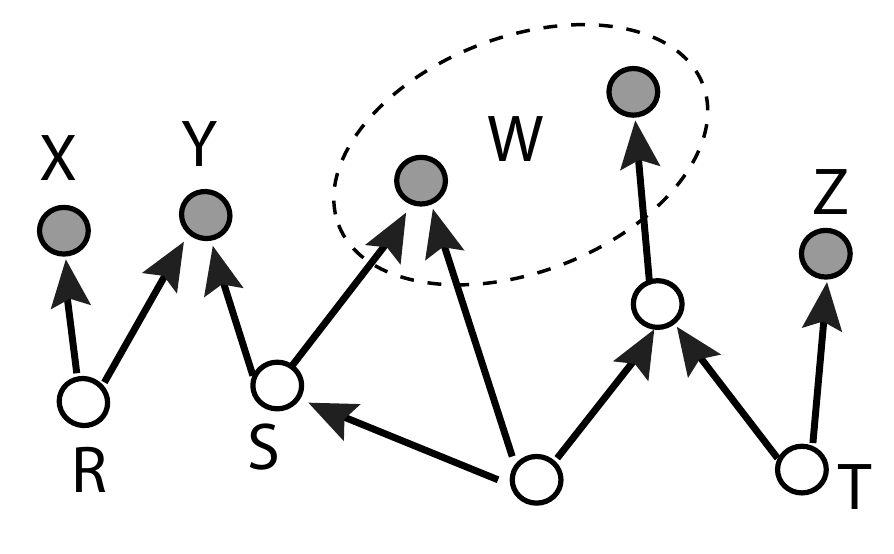}
\caption{ ``Chained" variables: $S$ and $Y$ are each connected to $T$ and $Z$ by a \tit{W-chain}. However $X$ and $R$ are $W$\tit{-detached} from $\{Z,T\}$, because the collider $Y$ is not in the set $W$. }
\label{fig:ochain}
\end{figure} 

(Note that when $O'$ is the empty set, the detached variables $\trm{\tbf{dt}}_{\emptyset}(V)$ are just those outcomes that do not share an ancestor with outcomes in $V$). Intuitively, if two variables in a DAG are connected by an $O'$-chain, they are connected by a path on which every collider has a directed path to $O'$. Hence, the detached variables are those nodes in the graph for which every path contains at least one collider that does \tit{not} lead to $O'$. This will be useful later when we consider graph separation. 

Let $\neg X$ denote the complement of a set $X$, and let $\neg_S X$ be the complement of $X$ restricted to $S$, i.e. $\neg_S X := S \cap \neg X$. Under a choice of ordering $\mathcal{O}$, let $S(<X_i)$ denote the set of predecessors of $X_i$ in $S$. Using these definitions, we propose the following characterisation of the CI relations in a quantum network:\\

\tbf{Definition \defin{def:qlist} \ref{def:qlist}: quantum input list.} A \tit{quantum input list} $Q_{\mathcal{O}}$ is a pair $\{ \trm{PA}_\mathcal{O}, Q \}$, containing:\\
\noindent \tbf{i.} An ordered list of parents, $\trm{PA}_\mathcal{O} := \{ \trm{\tbf{pa}}(X_i) : i=1,2,...,N \}$, where each set of parents $\trm{\tbf{pa}}(X_i)$ is a subset of $S(<X_i)$. Members of $O$ cannot be parents; in addition, every setting $S_i$ must be a parent of at least one other variable (these conventions ensure that the resulting DAG can be interpreted as a quantum circuit). Ancestors, descendants, etc, are defined from the list of parents in the usual way.\\
\tbf{ii.} A set of CI relations denoted $Q$, constructed as follows. For every subset of settings $S'$ and outcomes $O'$, there is a CI relation of the form $(S' \, \LL \, \trm{\tbf{dt}}_{O'}(S') | O' )$ and a CI relation of the form $(O'  \, \trm{\tbf{an}}(O') \LL \neg_S  \trm{\tbf{an}}(O' ) \, \trm{\tbf{dt}}_{\emptyset}(O' ) | \emptyset )$ in $Q$.  \\

 The first CI relation in the above definition expresses the physical requirement of setting independence, modulo the possibility of correlating the settings by conditioning on their effects. In particular, it says that settings are guaranteed independent except when connected by an $O'$-chain. The second CI relation simply expresses the Quantum Causality Condition. 

The quantum input list $Q_{\mathcal{O}}$ is said to be \tit{compatible} with a given probability distribution $P$ iff $\bar{Q} \subseteq \bar{C}(P)$. Given a quantum input list, we can construct a DAG in the usual way, by drawing a directed edge to each variable from each of its parent nodes. The DAG $G_Q$ constructed in this way is said to be \tit{generated} by the list $Q_{\mathcal{O}}$. As usual, the ancestors of $X$ are those nodes in the graph that have a directed path leading to $X$. The quantum input list defines the causal constraints on the variables, based on their interpretation as settings and outcomes in a quantum network. We conjecture that this list captures all of the conditional independencies that hold in a general quantum circuit, when the circuit is expressed as a DAG as outlined in Sec \ref{Sec:Qprelims}:\\

\tit{Conjecture:} If a CI relation holds in every quantum network represented by a DAG $G$, then it is implied by $Q$ in any quantum input list $Q_{\mathcal{O}}$ that generates $G$.\\

 \noindent So far, we have only specified the causal structure and independence relations. To obtain a full joint probability distribution from a quantum input list, we need to supplement it with model parameters specifying the pure state preparations, unitary transformations, and measurements that correspond to the variables. These parameters define the space of possible quantum circuits that are described by a given DAG:\\

\tbf{Definition \defin{def:qmodpar} \ref{def:qmodpar}: quantum model parameters.} Consider a set of variables $X_i$ with outcome spaces $\mathcal{E}_{X_i}$ connected in a DAG $G_Q$ representing a quantum network. Then the \tit{quantum model parameters} $F_q$ consist of:\\
\tbf{i.} A Hilbert space of dimension $2^n$ for each edge, where $n$ is a (possibly different) positive integer for each edge;\\
\tbf{ii.} A specification of the orthonormal basis $\mathcal{C}$ in which all projective measurements are made; \\
\tbf{iii.} For every exogenous node $X_i$, a pure state for every value in $\mathcal{E}_{X_i}$; \\
\tbf{iv.} For every intermediate node $X_i$, a unitary for every value in $\mathcal{E}_{X_i}$; \\
\tbf{v.} For every drain node $X_i$, a pure state from $\mathcal{C}$ for every value in $\mathcal{E}_{X_i}$; \\
\tbf{vi.} A marginal probability distribution on the outcome space $\mathcal{E}_{X_i}$ of every variable that is an exogenous node or intermediate node. These marginal distributions are all mutually independent.\\ 

The states and operators mentioned above apply to the Hilbert spaces of their respective nodes, as determined using \tbf{i} and the number of outgoing and ingoing edges (Recall that the dimensions assigned to the edges is constrained such that the total dimension of ingoing and outgoing edges for intermediate nodes is the same). The distributions given in \tbf{vi} represent a set of `initial conditions', fixed by the experimenters' choices and/or environmental conditions. These are used to determine the resulting probability distributions of the outcome variables according to the usual laws of quantum mechanics. This is made precise using the following definition:\\

\tbf{Definition \defin{def:QCM} \ref{def:QCM}: Quantum Causal Model}\\
A \tit{quantum causal model} (QCM) on a set of variables $X$ is a pair $\{Q_{\mathcal{O}}, F_q \}$ consisting of a quantum input list $Q_{\mathcal{O}}$ for the set $X$, and a set of quantum model parameters $F_q$ for the DAG $G_Q$ generated by the input list. \\

Every QCM defines a joint probability distribution over its variables according to the following procedure. Consider a QCM on $N$ ordered variables $\{ X_i : i=1,2,...,N \}$, and partition of the set $\{1,2,...,N \}:= \trm{E} \cup \trm{T} \cup \trm{D}$ such that $i \in \trm{E}$ labels the exogenous variables, $i \in \trm{T}$ labels the intermediate variables, and $i \in \trm{D}$ labels the variables corresponding to drains in the DAG $G_Q$. From $F_q$ we obtain the mutually independent marginal distributions $P(X_i) \, \forall i \in \trm{E} \cup \trm{T}$, which includes all setting variables. The joint probability of the outcomes conditional on the settings, $P( \cup_{i \in \trm{D} } X_i |  \cup_{i \in \trm{E},\trm{T}} X_i )$, is computed in the usual way from the quantum circuit obtained from the DAG $G_Q$ and the pure states and unitaries associated with the settings $ \{  \cup_{i \in \trm{E},\trm{T}} X_i  \} $. One thus obtains the total joint probability:
\eqn{
P(X) &=& \prod_{i \in \trm{E},\trm{T}} \, P(X_i) \,\, P( \cup_{j \in \trm{D}} X_j |  \cup_{j \in \trm{E},\trm{T}} X_j ) \, .
}

\tit{Note:} Our definition of a QCM on a DAG $G$ can be regarded as a concrete example of the more general notion of a \tit{Quantum Correlation} on the graph $G$, as defined by Fritz \cite{FRITZ}. In particular, working in the category $\mathcal{C}$ of completely positive maps, where Hilbert spaces are the objects and CP maps are the morphisms, we assign Hilbert spaces to the edges of the graph and CP maps to the nodes. The model parameters are just the set of functions from outcomes to morphisms that define the $\mathcal{C}$\tit{-instruments} in the language of Ref. \cite{FRITZ}, allowing us to compute probabilities. Our model distinguishes these functions according to the placement of their nodes in the graph (exogenous, intermediate or drain); we also model hidden variables as additional nodes, not as edges. These conventions do not represent limitations of our model, but are used for convenience. By contrast, our restriction to the case of variables with finite outcome spaces \tit{is} a limitation of our model, but we expect the generalisation to continuous variables following Ref. \cite{FRITZ} to be straightforward. 

Now that we have defined a QCM, we would like to have a graph separation rule analogous to d-separation that would allow us to recover all the CI relations implied by $Q_{\mathcal{O}}$ from the DAG $G_Q$. This is proposed in the next section.

\subsection{Graph separation in quantum networks \label{Sec:qsep}}

In general, because of the failure of the RCCP, we can never guarantee that two outcomes will be independent conditional on their common causes. However, there are still situations in which variables are expected to be conditionally independent of each other; we examine the possibilities below.

Two settings are already assumed to be chosen independently, so they can only become dependent on each other by conditioning on a common effect (which is an outcome), or conditioning on a connected chain of such effects. This applies also to conditioning on common effects in a CCM (recall Sec. \ref{Sec:CCMphysical}). In the case of a setting and an outcome, these might be dependent on each other if the setting is already a possible cause of the outcome, since the causal influence cannot in general be screened-off by other variables. On the other hand, if there is no directed path from the setting to the outcome and no chain of conditioned effects, one would expect the two to be independent. We must be careful, however: in quantum mechanics, it is also possible for the outcome to be entangled to another outcome that is descended from the setting, such that conditioning on the latter outcome correlates the setting with the causally separated outcome. To ensure their independence, therefore, one should not also not condition on any outcomes that are descended from the setting. Finally, two outcomes should be independent unless they share a common cause, or are connected by a chain of conditioned effects. 

These considerations lead us to the following graph separation criterion:\\

\tbf{Definition \defin{def:qsep} \ref{def:qsep}: q-separation}\\
Given a DAG representing a quantum network, two disjoint sets of variables $X$ and $Y$ are said to be \tit{q-separated} by a third disjoint set $Z$, denoted $(X \LL Y | Z)_q$, iff every undirected path between $X$ and $Y$ is rendered inactive by a member of $Z$. A path connecting two variables is rendered inactive by $Z$ iff \tit{at least} one of the following conditions is met:\\
(i) both variables are settings, and at least one of the settings has no directed path to any outcome in $Z$; \\
(ii) one variable is a setting and the other is an outcome, and there is no directed path from the setting to the outcome, or to any outcome in $Z$; \\
(iii) the path contains a collider $i \rightarrow m \leftarrow j$ where $m$ is not an outcome in $Z$, and there is no directed path from $m$ to any outcome in $Z$. \\

\begin{figure}[!htbp]
\includegraphics[width=8cm]{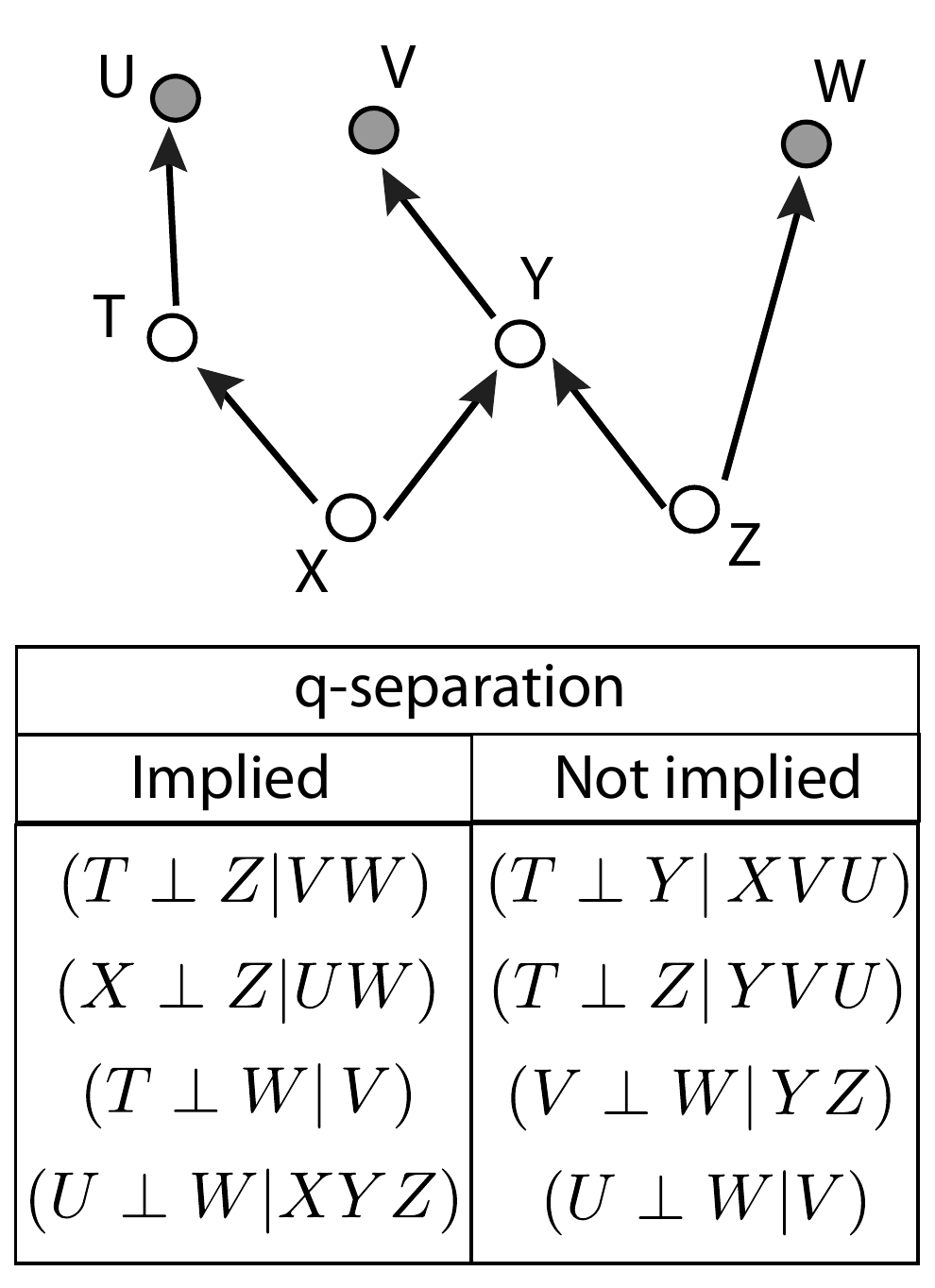}
\caption{An illustration of q-separation. The table contains CI relations that are either implied or not implied by the DAG using the rules of q-separation. Note the differences to d-separation (Fig. \ref{fig:dsep}).}
\label{fig:qseps}
\end{figure} 

Of course, the heuristic motivation given above does not necessarily guarantee that q-separation captures all of the CI relations that are implied by a quantum input list, nor is it obvious that the input list contains all CI relations implied by q-separation. A proof that q-separation is sound and complete for quantum input lists is given in the next section. 

\subsection{The q-separation theorem \label{Sec:qsepthrm}}

In this section we prove the soundness and completeness of q-separation. The proof approximately follows that of Pearl \& Verma \cite{VERMA} for the classical case. By analogy with d-separation, we will consider the set of CI relations obtainable from a DAG $G$ using the q-separation criterion (Definition \ref{def:qsep}) and $(X \LL Y| Z)_q \Rightarrow (X \LL Y | Z) $. Let this set be denoted $C_q(G)$, with $\bar{C}_q(G)$ its closure. If we replace d-separation with q-separation in Definition \ref{def:indep}, we obtain analogous criteria for $G$ to be an I-map or a perfect map of a given distribution $P$. We can now prove the following useful theorem:\\

\tbf{Theorem \thrm{thrm:qexist} \ref{thrm:qexist}: }\\
Let the DAG $G$ be a perfect map of a distribution $P(X)$ under q-separation, i.e. $\bar{C}_q(G)=\bar{C}(P)$. Then there is a quantum input list $Q_{\mathcal{O}}$ compatible with $P$ that generates the DAG $G$.\\

\tbf{Proof:} The DAG $G$ imposes a partial order on the variables $X$. Choose any total order $\mathcal{O}$ that is consistent with this. Label the nodes in $G$ as outcomes $O_i \in O$ if they are drains, and settings $S_i \in S$ otherwise. Define the parents $\trm{\tbf{pa}}(X_i)$ in $\trm{PA}_\mathcal{O}$ to be the nodes with directed edges pointing to $X_i$ in the graph. A path between a setting and any other variable is rendered inactive by outcomes $O'$ if there is at least one collider on the path not in $O'$ and with no directed path to $O'$. This is true for all variables that are $O'$-detached from the setting, hence the CI relation $(S' \, \LL \, \trm{\tbf{dt}}_{O'}(S') | O' )$ is implied by $G$. For each set of outcome nodes and their ancestors, $O' \trm{\tbf{an}}(O')$, a path from this set to the non-ancestors $\neg_S \trm{\tbf{an}}(O')$ can only be activated by conditioning on an outcome. Furthermore, a path from $O' \trm{\tbf{an}}(O')$ to $\trm{\tbf{dt}}_{\emptyset}(O')$ must contain a collider, or else $O'$ and $\trm{\tbf{dt}}_{\emptyset}(O')$ would share an ancestor (a contradiction), so it too can only be activated by conditioning on an outcome. Hence these paths are rendered inactive by the empty set, and the CI relation $(O' \trm{\tbf{an}}(O') \LL \neg_S \trm{\tbf{an}}(O') \, \trm{\tbf{dt}}_{\emptyset}(O') | \emptyset)$ is implied by $G$. According to Definition \ref{def:qlist}, these ingredients are sufficient to specify a quantum input list. By construction, this list also generates the DAG $G$. $\Box$ \\

The next theorem provides the key result.\\

\tbf{Theorem \thrm{thrm:qsep} \ref{thrm:qsep}}\\
Given a distribution $P(X)$ and a compatible quantum input list $Q_{\mathcal{O}}$, the DAG $G$ generated by $Q_{\mathcal{O}}$ is an I-map of $P$, that is, $\bar{C}_q(G) \subseteq \bar{C}(P)$.\\

\tit{Proof:} We prove the result by induction on the number of variables. First we show that the result holds for $k$ variables, given that it holds for $k-1$ variables. Then we note that the result holds trivially for one variable; hence, by induction, it holds for any number of variables.

Let $P$ be a distribution on $k$ variables and $Q_{\mathcal{O}}$ a compatible quantum input list, which generates the DAG $G$. Let $n$ be the last variable in the ordering $\mathcal{O}$; let $\bar{C}(P-n)$ be the closed set of CI relations formed after removing from $\bar{C}(P)$ all CI relations involving $n$; let $P-n$ be any probability distribution having exactly the closed set of CI relations $\bar{C}(P-n)$ (such a distribution can always be constructed \cite{PEARL1}); and let $G-n$ be the DAG formed by removing the node $n$ and all its connected edges from the graph $G$.

Consider the list obtained from $Q_{\mathcal{O}}$ by removing every CI relation involving $n$ from $Q$ and removing $\trm{\tbf{pa}}(n)$ from the list of parents (let the reduced list of parents be denoted $\trm{PA}_\mathcal{O}-n$). This procedure might result in one or more settings that are not parents of any other variables, so we must be convert these into outcomes in order to make a valid input list. To do so, we first remove the setting in question (say $S_j$) from any sets of settings $S'$ in which it appears in $Q$. Next, we re-label it as an outcome, $S_j \rightarrow O_j$. This entails that for every subset $O'$ of outcomes (not containing $n$) and settings $S'$, we must add new CI relations $( O_j O' \, \trm{\tbf{an}}( O_j O') \LL \neg_S \trm{\tbf{an}}( O_j O') \trm{\tbf{dt}}_{\emptyset}( O_j O') | \emptyset)$ and $(S' \, \LL \, \trm{\tbf{dt}}_{O_j O'}(S') | O_j O' )$ to $Q$. Let the resulting list be denoted $Q-n$. Then the pair $Q_{\mathcal{O}}-n := \{\trm{PA}_\mathcal{O}-n, Q-n \}$ is a valid quantum input list on $k-1$ variables. Furthermore, by construction, $Q_{\mathcal{O}}-n$ generates the DAG $G-n$. 

Let us now assume that $G-n$ is an I-map of $P-n$: $\bar{C}_q(G-n) \subseteq \bar{C}(P-n)$. We aim to prove that, under this assumption, $G$ is also an I-map of $P$,  $\bar{C}_q(G) \subseteq \bar{C}(P)$. To do so, we will consider each CI relation of $\bar{C}_q(G)$ and show that it exists also in $\bar{C}(P)$. 

The CI relations of $\bar{C}_q(G)$ can be divided into three cases: \\

\noindent (1) $n$ does not appear in the CI relation; \\
(2) $n$ appears in the first position in the CI relation, eg. $(nX \LL Y| Z)$; \\
(3) $n$ appears in the last position in the CI relation, eg. $(X \LL Y| nZ)$.\\

Note that, if $n$ appears in the second position in the CI relation, we can use symmetry (semi-graphoid axiom 1.a.) to move it into the first position and thereby convert it into case (2) above. We now prove the result for each case separately.\\

\noindent \tbf{Lemma 1:} Let $X,Y,Z$ be disjoint sets of variables and let $R \in \bar{C}_q(G)$ be a relation of the form $(X \LL Y | Z)$ that does not contain the variable $n$. Then $R \in \bar{C}(P)$.\\ 

\tit{Proof:} Since $R$ is in $\bar{C}_q(G)$, it must also be in $\bar{C}_q(G-n)$. If it were not, then there would be a path between $X$ and $Y$ that is active in $\bar{C}_q(G-n)$ but rendered inactive by $Z$ in $\bar{C}_q(G)$. But this is impossible, because an active path cannot be rendered inactive just by adding a node and its associated edges to the graph. Since $G-n$ is an I-map of $P-n$, the relation $R$ must also be contained in $\bar{C}(P-n)$, and since $\bar{C}(P-n)$ is a subset of $\bar{C}(P)$, $R$ is also contained in $\bar{C}(P)$. $\Box$ \\

\noindent \tbf{Lemma 2:} Let $R \in \bar{C}_q(G)$ be a relation of the form $(nX \LL Y| Z)$. Then $R \in \bar{C}(P)$. \\

\tit{Proof:} First, we partition the sets $X,Y,Z$ into disjoint sets of outcomes and settings, eg. $Z=Z_O \cup Z_S$ where $Z_O$ contains only outcomes and $Z_S$ only settings. 

 Define the set $\trm{\tbf{zo}}(O')$ as the members of $Z_O$ that are connected to outcomes $O'$ by a $Z_O$-chain. Let $\trm{\tbf{-zo}}(O')$ denote its complement in $Z_O$. Next, consider $O_x:=n X_O \trm{\tbf{zo}}(n X_O)$.  

\begin{figure}[!htbp]
\includegraphics[width=8cm]{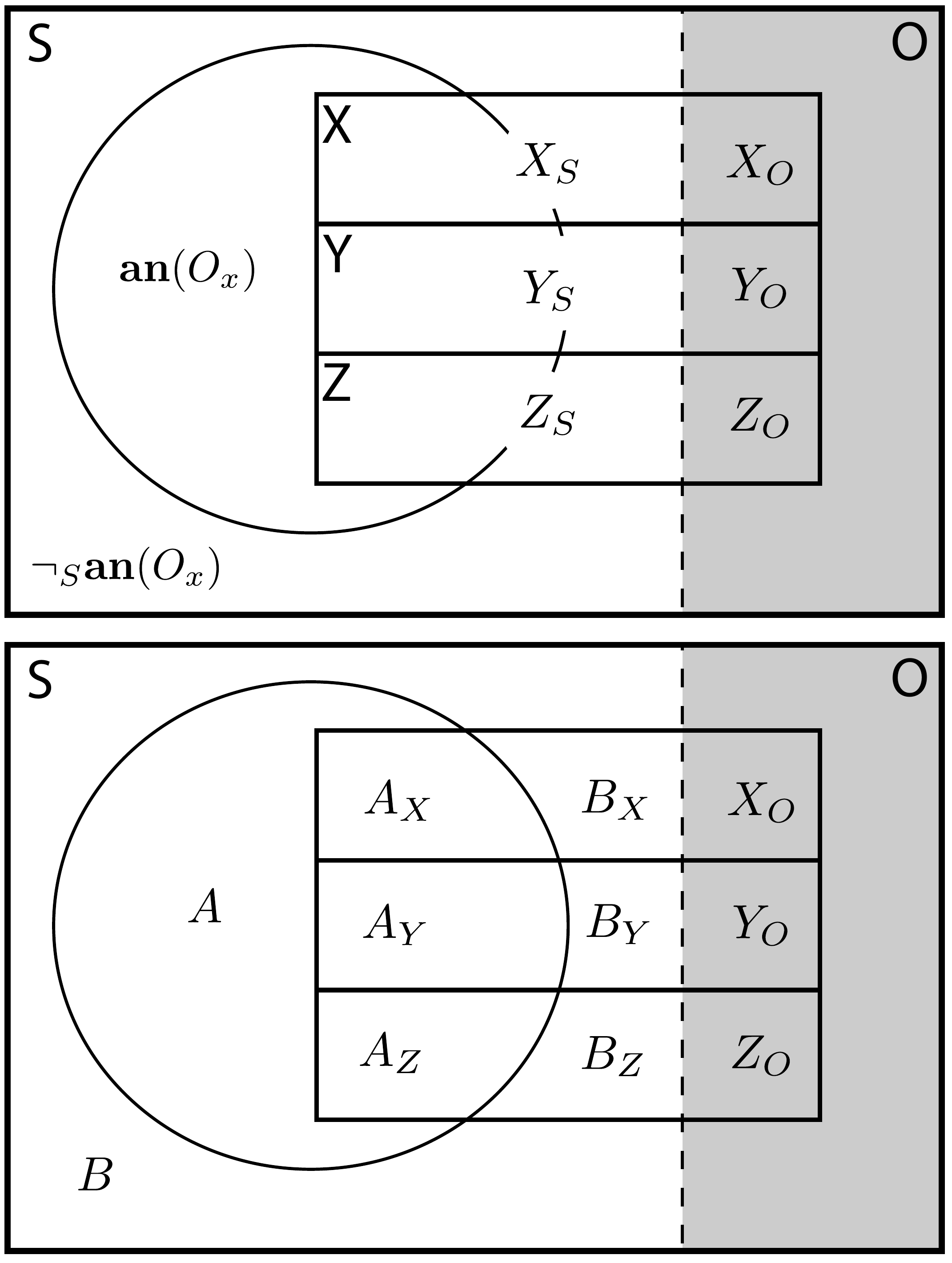}
\caption{A partitioning of the variables into disjoint sets. Above: the sets $X,Y,Z$ are partitioned with respect to their settings, outcomes and $\trm{\tbf{an}}( O_x )$. Below: the ancestors and their complement in $S$ are further decomposed into disjoint subsets.}
\label{fig:Venn}
\end{figure}

We can write the ancestors $\trm{\tbf{an}}( O_x ))$ as the union of four disjoint sets, $\trm{\tbf{an}}( O_x ) = A_X \cup A_Y \cup A_Z \cup A$, where $A_X := \trm{\tbf{an}}( O_x ) \cap X$, and similarly for $A_Y$ and $A_Z$. Any remaining members of $\trm{\tbf{an}}( O_x )$ not contained in any of $X,Y,Z$ are contained in $A$. Likewise, let us decompose $\neg_S \trm{\tbf{an}}( O_x)$ into disjoint sets: $\neg_S \trm{\tbf{an}}( O_x) = B_X \cup B_Y \cup B_Z \cup B$ where eg. $B_X := \neg_S \trm{\tbf{an}}( O_x) \cap X$ and similarly for $B_Y, B_Z$ (see Fig. \ref{fig:Venn}). Note that $X_S = A_X \cup B_X$ and analogously for $Y$ and $Z$. The CI relation $(O_x \, \trm{\tbf{an}}(O_x) \LL \neg_S \trm{\tbf{an}}(O_x) \trm{\tbf{dt}}_{\emptyset}(O_x) | \emptyset)$ must hold in $Q$ and hence in $\bar{C}(P)$. Using the above definitions: 
\eqn{  \label{lem2eq0}
&& ( O_x \, \trm{\tbf{an}}( O_x) \LL \neg_S \trm{\tbf{an}} ( O_x)  \, \trm{\tbf{dt}}_{\emptyset} (O_x) | \emptyset) \in \bar{C}(P) \nonumber \\
& \implies & ( O_x  \LL \neg_S \trm{\tbf{an}} ( O_x)  \, \trm{\tbf{dt}}_{\emptyset} ( O_x)  | \trm{\tbf{an}}( O_x)) \in \bar{C}(P) \nonumber \\
& \Longleftrightarrow &  ( O_x  \LL B_X  B_Y  B_Z  B  \, \trm{\tbf{dt}}_{\emptyset} ( O_x) |  A_X A_Y A_Z A) \in \bar{C}(P) \, . \nonumber \\
&&
}

 The set $\trm{\tbf{-zo}} (n X_O)$ must be a subset of $\trm{\tbf{dt}}_{\emptyset} ( O_x)$ (since the members of $\trm{\tbf{-zo}} (n X_O)$ by definition cannot share an ancestor with $O_x$). The same goes for $Y_O$, otherwise there would be a path connecting $Y_O$ to $n X_O$ on which every collider is in $Z_O$ or has a directed path to $Z_O$, and they could not be q-separated in $G$ as is required for $R$ to be true. Hence, using the semi-graphoid axioms:
\eqn{  \label{lem2eq1}
&& ( O_x \LL B_X  B_Y  B_Z  B \, \trm{\tbf{dt}}_{\emptyset} ( O_x) |  A_X A_Y A_Z A) \in \bar{C}(P) \nonumber \\
&\implies& ( O_x \LL  B_Y  B_Z  \, Y_O \, \trm{\tbf{-zo}}(n X_O) | X_S A_Y A_Z ) \in \bar{C}(P) \nonumber \\
&\implies& ( n  \LL B_Y Y_O |  X Z A_Y) \in \bar{C}(P) \, . 
}
 No member of $Y$ can be an ancestor of $n X_O$ in $Q$, or else there would be a directed path from a setting in $Y_S$ to an outcome in $n X$ and they could not be q-separated by $Z$ in $G$, contradicting our initial premise $R$. Therefore $A_Y=\emptyset $ and $B_Y=Y_S $, and \eqref{lem2eq1} implies $( n  \LL Y |  X Z) \in \bar{C}(P)$. The relation $R$ implies $(X \LL Y| Z) \in \bar{C}_q(G)$ and hence (by Lemma 1): $(X \LL Y| Z) \in \bar{C}(P)$. Combining this with \eqref{lem2eq1} and the semi-graphoid axioms, we obtain the desired result $(n X \LL Y| Z ) \in \bar{C}(P)$. $\Box$ \\

\noindent \tbf{Lemma 3:} Let $R \in \bar{C}_q(G)$ be a relation of the form $(X \LL Y| nZ)$. Then $R \in \bar{C}(P)$. \\

\tit{Proof:}  Note that $n$ cannot share an ancestor with both $X_O \trm{\tbf{zo}}(X_O)$ \tit{and} $Y_O \trm{\tbf{zo}}(Y_O)$, or else there would be a path connecting $X_O$ to $Y_O$ on which every collider has a descendant in $n Z_O$, preventing them from being q-separated in $G$. We therefore assume without loss of generality that $n$ does not share an ancestor with $Y_O \trm{\tbf{zo}}(Y_O)$. By a similar argument, no member of $X_O$ can share an ancestor with $Y_O \trm{\tbf{zo}}(Y_O)$; hence $n X_O \trm{\tbf{-zo}}(Y_O) \in \trm{\tbf{dt}}_{\emptyset} (Y_O \trm{\tbf{zo}}(Y_O))$. Now, either $n$ shares an ancestor with $X_O \trm{\tbf{zo}}(X_O)$, or it does not. If it does, then $Y$ cannot contain any ancestors of $O_x$ (defined as in the previous Lemma) and we can use the same procedure as before to obtain the desired result: $(n X \LL Y| Z ) \in \bar{C}(P)$. 

In the remaining case, $n$ does not share an ancestor with $X_O \trm{\tbf{zo}}(X_O)$ so we have the relation $n Y_O \trm{\tbf{-zo}}(X_O) \in \trm{\tbf{dt}}_{\emptyset}(X_O \trm{\tbf{zo}}(X_O))$. Let $O_{xz}:=X_O \trm{\tbf{zo}}(X_O)$ and consider the relation $( O_{xz} \, \trm{\tbf{an}}(O_{xz}) \LL \neg_S \trm{\tbf{an}}(O_{xz}) \, \trm{\tbf{dt}}_{\emptyset}(O_{xz}) | \emptyset)$ that holds in $Q$ and hence in $\bar{C}(P)$. Using the above properties, and the fact that $Y$ cannot contain any ancestors of $O_{xz}$ (for the usual reason that this would imply an active path between $X$ and $Y$ in $G$), we obtain:
\eqn{ \label{lem3eq1}
&& ( O_{xz} \, \trm{\tbf{an}}(O_{xz}) \LL \neg_S \trm{\tbf{an}}(O_{xz}) \, \trm{\tbf{dt}}_{\emptyset}(O_{xz}) | \emptyset) \in \bar{C}(P) \nonumber \\
&\implies& ( X_O \, \trm{\tbf{zo}}(X_O)  \LL Y_S \, n Y_O \, \trm{\tbf{-zo}}(X_O) | X_S Z_S ) \in \bar{C}(P) \nonumber \\
&\implies& ( X_O  \LL  Y | n X_S \, Z ) \in \bar{C}(P) \, .
}
Let us partition $Z_S= D_Z \cap E_Z$, where $D_Z$ contains the members of $Z_S$ that are detached from $X_S$ by $n Z_O$, and $E_Z$ contains the rest. Consider the CI relation $( X_S E_Z \LL \trm{\tbf{dt}}_{n Z_O}(X_S E_Z) | n Z_O)$ that holds in $Q$ and hence in $\bar{C}(P)$. The set $\trm{\tbf{dt}}_{n Z_O}(X_S E_Z)$ must contain $\trm{\tbf{dt}}_{n Z_O}(X_S)$. If not, there would be a member of $\trm{\tbf{dt}}_{n Z_O}(X_S)$ that is not detached from $X_S E_Z$, hence it would be connected by an $n Z_O$-chain to $E_Z$. But since $E_Z$ has a chain to $X_S$, it could not be a member of $\trm{\tbf{dt}}_{n Z_O}(X_S)$, implying a contradiction. Hence $( X_S E_Z \LL \trm{\tbf{dt}}_{n Z_O}(X_S) | n Z_O) \in \bar{C}(P)$. But note that $Y$ must be detached from $X_S$ by $n Z_O$, otherwise there would be a path connecting $X$ to $Y$ in $G$ on which every collider has a directed path to $n Z_O$, contradicting $R$. Thus $D_Z Y \in \trm{\tbf{dt}}_{n Z_O}(X_S)$, and we obtain:
\eqn{ \label{lem3eq2}
&& ( X_S E_Z \LL \trm{\tbf{dt}}_{n Z_O}(X_S) | n Z_O) \in \bar{C}(P) \nonumber \\
&\implies& ( X_S \LL Y | n Z) \in \bar{C}(P) \, .
}
Combining this result with \eqref{lem3eq1} and the Contraction axiom 1.d, we obtain $( X  \LL  Y | n Z ) \in \bar{C}(P)$ as desired.  $\Box$ \\

Lemmas 1,2 and 3 together imply that $G$ is an I-map of $P$, i.e. $\bar{C}_q(G) \subseteq \bar{C}(P)$, provided that $G-n$ is an I-map of $P-n$. The latter condition can be guaranteed using the same logic: $G-n$ is an I-map of $P-n$ provided $G-n-m$ is an I-map of $P-n-m$, where $m$ is now the second-last variable in the chosen ordering. Continuing this process, every graph in the hierarchy is an I-map of its corresponding distribution provided that we can prove that $G$ for a single variable is an I-map of a probability distribution $P$ on a single variable. But $\bar{C}_q(G) \subseteq  \bar{C}(P)$ is trivially satisfied for a single variable, because both these sets are empty. This completes the proof of Theorem \ref{thrm:qsep}. $\Box$ \\

Using the result of Theorem \ref{thrm:qsep}, we can now prove that q-separation is sound and complete for quantum input lists.\\

\tbf{Theorem \thrm{thrm:corol} \ref{thrm:corol}}\\
If $Q$ is a quantum input list, the DAG generated by $Q$ is a perfect map of its semi-graphoid closure $\bar{Q}$. That is, a CI relation follows from the DAG if and only if it can be obtained from $Q$ using the semi-graphoid axioms.\\

\tit{Proof:} By the previous theorem, the DAG is an I-map of $\bar{Q}$. It remains to show the converse, namely that every relation in $\bar{Q}$ is implied by the DAG: $\bar{Q} \subseteq \bar{C}_q(G)$. Since $\bar{C}_q(G)$ is a semi-graphoid, it contains the closure of $\bar{Q}$ if it contains $Q$. Hence it suffices to show $Q \subseteq \bar{C}_q(G)$. Consider any CI relation of the form $( S' \LL \trm{\tbf{dt}}_{O'}(S') |O') \in Q$. Since $Q$ generates $G$, the set $\trm{\tbf{dt}}_{O'}(S')$ consists of all nodes whose paths to $S'$ contain at least one collider that is not in $O'$ and has no directed path to $O'$. This is sufficient for these paths to be rendered inactive by $O'$, hence $( S' \LL \trm{\tbf{dt}}_{O'}(S') |O')_q \in G$. Next consider $( O' \, \trm{\tbf{an}}(O') \LL \neg_S \trm{\tbf{an}}(O') \, \trm{\tbf{dt}}_{\emptyset}(O') | \emptyset) \in Q$. Since $\trm{\tbf{dt}}_{\emptyset}(O')$ shares no ancestors with $O'$, any path between them must contain at least one collider, hence be rendered inactive by the empty set. Since the empty set obviously does not contain any ancestors of $S$ and since $\trm{\tbf{an}}(O')$ has no directed path to $\trm{\tbf{dt}}_{\emptyset}(O')$ and $\neg_S \trm{\tbf{an}}(O') $ has no directed path to $O'$, we conclude that all paths between $O' \, \trm{\tbf{an}}(O') $ and $\neg_S \trm{\tbf{an}}(O') \, \trm{\tbf{dt}}_{\emptyset}(O')$ are inactive in $G$, so $( O' \, \trm{\tbf{an}}(O') \LL \neg_S \trm{\tbf{an}}(O') \, \trm{\tbf{dt}}_{\emptyset}(O') | \emptyset)_q \in G$. This covers all possible CI relations in $Q$, so $\bar{Q} \subseteq \bar{C}_q(G)$. $\Box$  \\

\subsection{Correspondence to classical models \label{Sec:Corresp}}

In a quantum circuit, one can obtain a `classical limit' by restricting all states and operators to a subspace of Hilbert space. In order to define the classical limit of a QCM, we must ensure that, in this limit, we recover the assumptions listed in Sec. \ref{Sec:CCM}. We expect that Reichenbach's Common Cause Principle will be recovered by restricting our operations to a classical subspace of Hilbert space, since this will rule out the possibility of entanglement. However, the Causal Markov Condition requires that direct causes `screen-off' indirect causes, which will not in general be true after restricting the circuit to a classical subspace. To recover this principle, therefore, we need to transform the DAG of the QCM into a form that respects this property. This can be done by assigning the ancestors of every outcome (the `possible causes' by Assumption \ref{ass:pastcone}) to either direct or indirect causes. There may be many ways to do this, so we will adopt the simplest solution and make them all direct causes. This procedure is described in the following definition:\\

\tbf{Definition \defin{def:classlim} \ref{def:classlim}: Classical limit for DAGs}
Given a DAG $G$ interpreted as a quantum network, the \tit{classical limit} of $G$ is a new DAG $G_C$ obtained by the following procedure:\\
1. Draw a directed edge from every setting $S_i$ (non-drain node) to every outcome (drain node) $O_i$ that is descended from $S_i$ in $G_C$, unless such an edge already exists.\\
2. Eliminate all edges that connect pairs of settings to each other.\\
 
The first step makes every setting a direct cause of every outcome that is descended from it, while the second step uses the fact that the settings are mutually independent to eliminate redundant edges. The resulting DAG is consistent with the causal structure of the original DAG in the following sense: there is a causal chain from one variable to another in $G_C$ \tit{only if} there existed such a chain in the DAG of the quantum network $G$. The the screening-off property is enforced since there are no intermediate nodes left in the DAG. This allows us to recover d-separation in the classical limit:\\

\tbf{Theorem \thrm{thrm:classlim} \ref{thrm:classlim}}\\
Given any DAG $G$ with classical limit $G_C$, every CI relation obtainable from $G_C$ by q-separation is also implied by d-separation, i.e. $\bar{C}_q(G_C) \subseteq \bar{C}_{d}(G_C)$. \\

\tit{Proof:} Suppose $(S_A \LL S_B | V)_q$ holds under q-separation for disjoint sets $S_A,S_B,V$ where $S_A, S_B$ are settings. Since all paths connecting two settings in $G_C$ must contain at least one collider that is an outcome, q-separation implies that at least one of these colliders is not in $V$, but this also implies that the settings are d-separated. Similarly, if $(S_A \LL O_B | V)_q$ holds under q-separation for a set of outcomes $O_B$, this implies that $O_B$ is not descended from $S_A$. It follows that every path between them must contain at least one collider, and at least one of these colliders is not in $V$, which also implies d-separation. Finally, if $(O_A \LL O_B | V)_q$ holds for two sets of outcomes, then every path connecting them must contain a collider that is not in $V$, again implying d-separation.  $\Box$ \\

Given this result, it is straightforward to convert a QCM into a CCM: we restrict the quantum model parameters to a classical subspace and obtain a classical circuit. This circuit defines a set of functions $F$ that determine the values of the outcomes given the values of their ancestors (these become their parents in $G_C$ obtained from $G_Q$). The pair $\{G_C, F \}$ then satisfies the requirements of a CCM.

\subsection{More general correlations \label{Sec:PRBox}}

We note that the ``Bell-type" experiment described in Sec. \ref{Sec:QMintro} also applies more generally to any joint probability distribution with settings and outcomes that obey the no-signalling criterion. In particular, one can find a joint probability distribution $P$ on the variables $A,B,S,T$ that satisfies the CI relations $K$ (implied by setting independence and no-signalling) but which cannot be generated by any QCM defined on the same variables. For example, let all variables be binary variables taking values in $\{0, 1\}$, let $\oplus$ represent addition modulo $2$, and consider the joint distribution:
\eqn{ \label{eqPRbox}
P(A,B,S,T) &:=& P(S)P(T)P(A,B | S,T) \, , \;\; \trm{where} \nonumber \\
 P(A,B | S,T) &=& \frac{1}{2} (1 \oplus A \oplus B) (ST \oplus 1) + \frac{1}{2} (A \oplus B) ST\, . \nonumber \\
 &&
}
One can check that this distribution satisfies the CI relations $K$ and that the probabilities sum to one. This distribution characterises a \tit{Popescu-Rorlich box} (PR-box), also called a nonlocal or non-signalling box \cite{PRBOX}. A PR-box defines correlations that are stronger than quantum correlations, in the sense that they violate Bell's inequality to its algebraic maximum. How do super-quantum correlations fit into the present framework of quantum causal models?

\begin{figure}[!htbp]
\includegraphics[width=8cm]{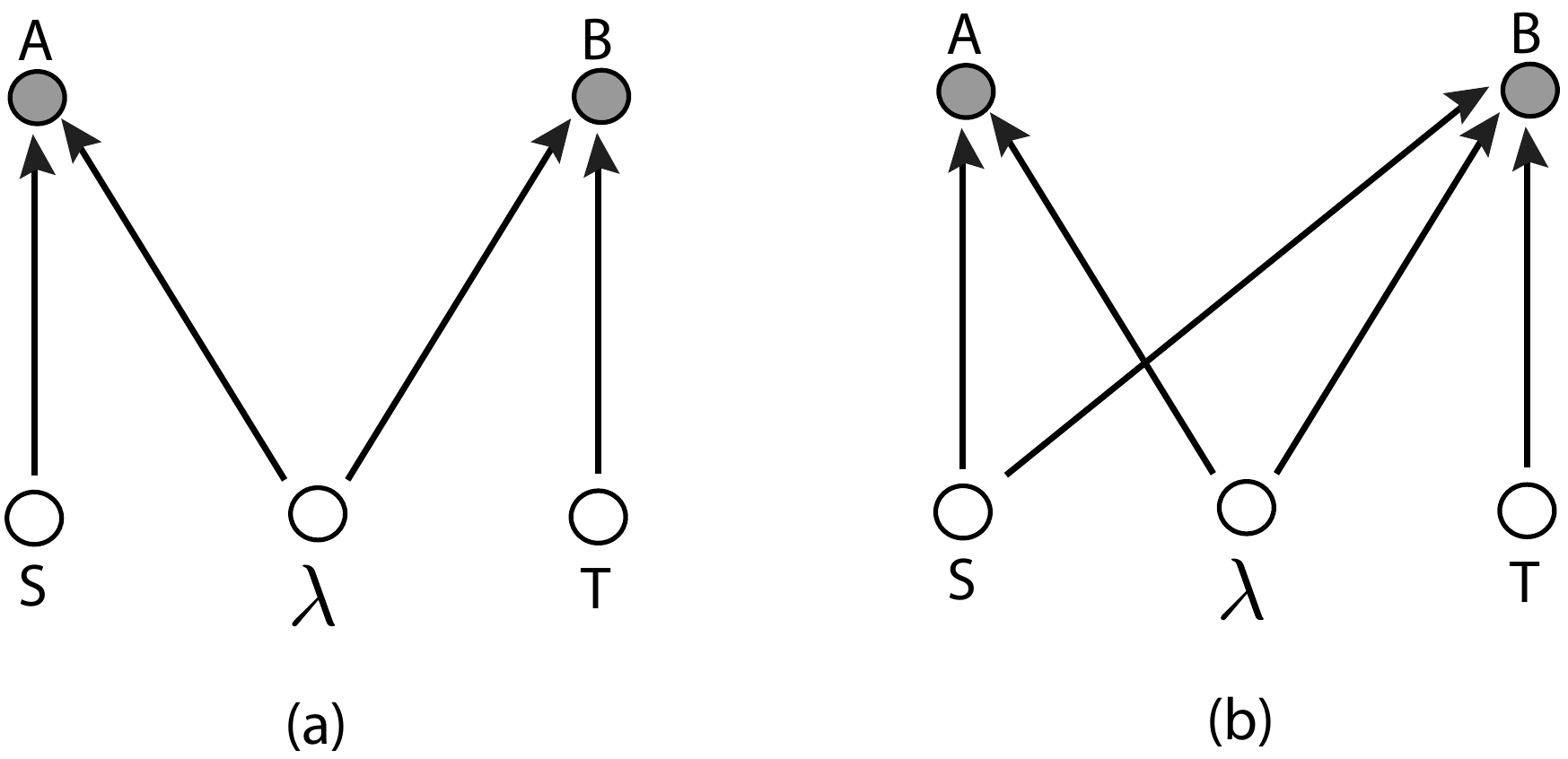}
\caption{(a) A DAG representing a quantum network that satisfies the CI relations of a PR-box, but cannot reproduce the full statistics. (b) A DAG representing a quantum network that can reproduce the PR-box, but is unfaithful because it requires fine-tuning.}
\label{fig:PRBell}
\end{figure}

There is no QCM on just the variables $A,B,S,T$ that provides a faithful explanation of the probability distribution $P$. This is trivially true because the constraints in $K$ require that there is no directed edge from $S$ to $B$ or from $T$ to $A$ in the corresponding DAG (or else they would not be q-separated by any subset of variables) and no outgoing edges from $A$ and $B$, which are outcomes - but this splits the DAG into two disconnected parts, which are necessarily independent, implying no correlations between $A$ and $B$. As before, we can try to explain the correlations by introducing a hidden variable $\lambda$ and extending the constraints to the set $K'$ (see Sec. \ref{Sec:QMintro}). For a QCM, this is equivalent to supposing that the outcomes $A$ and $B$ can depend on a shared entangled resource, in a pure state specified by $\lambda$; the corresponding DAG is shown in Fig. \ref{fig:PRBell} (a). It should therefore come as no surprise that a QCM based on this DAG fails to reproduce the distribution $P$. Whichever pure states one chooses for the values of $\lambda$, the resulting statistics obtained from the QCM must obey quantum mechanics and hence the joint probability generated by the QCM must be able to violate Bell's inequality only up to Tsirelson's bound.

Of course, there is no reason to restrict ourselves to this DAG -- just as in the classical case, we can consider DAGs such as the one shown in Fig. \ref{fig:PRBell} (b), in which there is a signal from $S$ to $B$, representing a `hidden' link in the underlying quantum network (\tit{Note:} If we interpret this DAG as a classical causal model, it also serves as an alternative to Fig. \ref{fig:PRBell} (b) for explaining quantum or PR-Box correlations using fine-tuning). This additional link can be used to send a single bit of information from $S$ to $B$, which can be exploited to simulate the desired probability distribution. But, just as in the classical case, this would imply that $S$ and $B$ should not be q-separated, so the CI relation $(S \LL B | T \lambda) \in \bar{K}'$ can only be due to fine-tuning of the model parameters and the explanation is not faithful. Just as a CCM cannot faithfully explain quantum correlations, a QCM also cannot faithfully explain super-quantum correlations. 

Despite the analogy, there is an important difference: the threshold between quantum and classical statistics is set by Bell's inequality, for which the only relevant constraints are the CI relations: if the statistics satisfy the CI relations $\bar{K}'$, then they satisfy Bell's inequality. The same is not true for the transition from quantum to super-quantum correlations. Both the statistics of the PR-box \eqref{eqPRbox}, and the statistics generated by the quantum network Fig. \ref{fig:PRBell} (a) satisfy the CI relations $\bar{K}'$. The difference lies in the constraints on the model parameters; while the model parameters of a CCM are unconstrained (except as dictated by the CI relations), the model parameters of a QCM are additionally constrained by Definition \ref{def:qmodpar}, which restricts all operations to the standard quantum formalism.

\section{Discussion and Conclusions \label{Sec:Conc}}

Given a probability distribution, a classical causal model (CCM) describes the causal connections of events compatible with the observed correlations under the assumption that these correlations are generated by classical physics. We have shown that the same can by done under the assumption that the underlying physics is quantum. We gave a suitable definition of a quantum causal model (QCM) on a directed acyclic graph that represents a quantum network and is consistent with related work, namely Ref. \cite{FRITZ}. We showed that it is possible to deduce the CI relations implied by a the quantum network using a criterion called q-separation, which we proved to be sound and complete. In principle, an algorithm based on q-separation could be used to program an artificial intelligence to make inferences about the connections between the components of a quantum network, given only the observed correlations between the variables in the network. It is left to future work to investigate whether such an algorithm presents any practical advantage over approaches based on CCMs. 

\begin{figure}[!htbp]
\includegraphics[width=8cm]{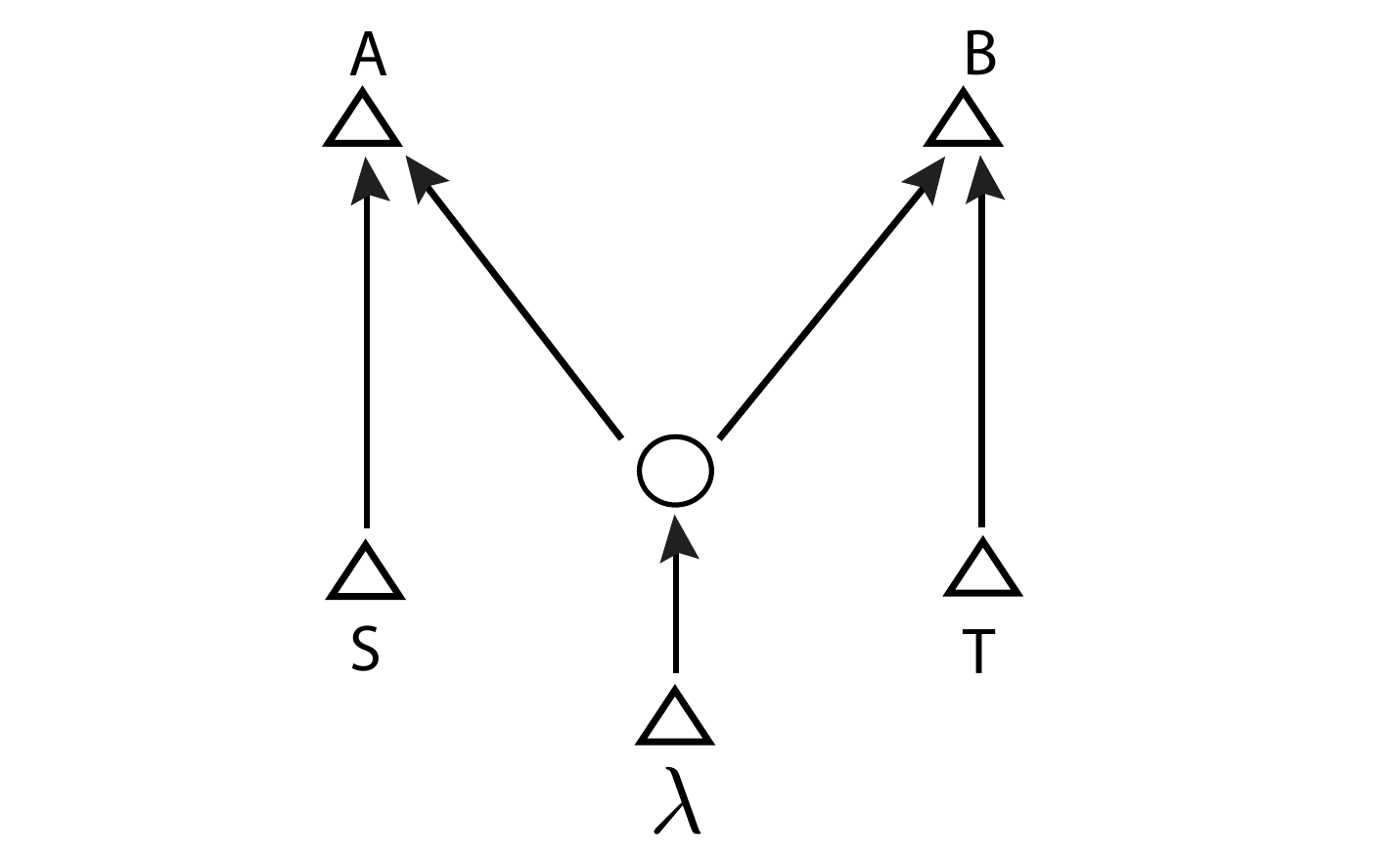}
\caption{An alternative representation of the Bell scenario due to Ref. \cite{HLP}, in which the correct CI relations for a shared quantum resource are obtained using only d-separation. The formalism relied on the introduction of an unobserved circular node that cannot be conditioned upon.}
\label{fig:hlpbell}
\end{figure}

It is interesting to compare the approach to graph separation taken here to that of Ref. \cite{HLP}, in which a DAG representation of generalised probabilistic theories was proposed that retains the d-separation rule. If we consider the Bell scenario in this latter framework, we obtain a graph like that of Fig. \ref{fig:hlpbell}, where now $\lambda$ represents the in-principle observable preparation of an entangled resource, as in our formalism, but there is an additional `unobserved' node, depicted as a circle, representing the quantum nature of the resource. The remaining nodes are depicted as triangles to indicate that they are observed. The presence of the unobserved node ensures that the two outcomes are no longer d-separated by any subset of the \tit{observed} variables $S,T,\lambda$. 

Of course, if it were possible to condition on the unobserved node, we would not have progressed from a CCM. Rather, the purpose of these unobserved nodes is to restrict us to a special subset of the CI relations obtainable by d-separation from the graph, which is then proven to be the correct set of CI relations for a quantum network (or a generalised probabilistic theory). A possible advantage of retaining d-separation in this way is that the existing algorithms for extracting conditional independencies from a graph still apply and can be used in a practical setting by a computer program. However, it could be argued that this approach misses something of the structure that underlies the CI relations in a quantum setting, which is made explicit in the present work through the definition of q-separation. Thus, the present work is complementary in that it elucidates those constraints on the causal structure of quantum networks that remain after the factorisation property of the RCCP is dropped. 

We have argued that there still exist non-trivial physical constraints on the conditional independencies between variables in a general quantum network, even without the RCCP. Our observation that super-quantum correlations are differentiated from quantum correlations only by their model parameters (and not by their CI relations) in the Bell scenario indicates that the same constraints may apply generally, and that they stem from the Markovianity condition; this is supported by the work of Ref. \cite{FRITZ}. Unfortunately, this also indicates that relaxations of the RCCP alone may not distinguish quantum theory from more general probabilistic theories. However, we have not proven that q-separation is sound and complete for a suitably chosen DAG formulation of any generalised probabilistic theory, so it is left as an open question, as is the question of whether one can generalise the RCCP to a principle that can distinguish quantum from general probabilistic theories, not just from classical causal models. 

Finally, we speculate that the type of causal model discussed here might present a starting point for investigations into the quantum nature of space-time. After all, if the space-time manifold of classical general relativity is to give way to a more fundamental structure at the Planck scale, then it seems plausible that this structure should consist of something like a causal network, which supplies the essential geometric information about some discrete set of fundamental events. It would be interesting to see whether such a construction could make connections to existing work on quantum gravity, such as spin foams and causal sets, and whether it can be generalised to include more exotic effects, such as closed time-like loops or the recently proposed phenomenon of `indefinite causality' \cite{ORE12,ORE14}.

\tit{Acknowledgements:} We thank M. Pusey for drawing our attention to a flaw in an earlier draft of the paper (now fixed). This work has been supported by the European Commission Project RAQUEL, the John Templeton Foundation, FQXi, and the Austrian Science Fund (FWF) through CoQuS, SFB FoQuS, and the Individual Project 2462.

\end{document}